\documentclass{aa}  
\usepackage{graphicx}   
\usepackage{amsmath}    
\usepackage{amssymb}    
\usepackage{multicol}        
\usepackage{booktabs}
\usepackage{multirow}
\usepackage{rotating}
\usepackage{siunitx}
\usepackage{bm}         
\usepackage{pdflscape}  

\usepackage{subcaption}

\usepackage{txfonts}
\usepackage[colorlinks=true, linkcolor=blue,anchorcolor=black,citecolor=blue,filecolor=black,menucolor=black,runcolor=black,urlcolor=black]{hyperref}
\usepackage[nameinlink,capitalise]{cleveref}

\newcommand{\mbh}{ M_{\rm BH}}

\newcommand{\Msun}{\rm M_{\sun}}

\newcommand{\mstar}{M_{\rm *}}

\newcommand{\Ledd}{\lambda_{\rm Edd}}
\newcommand{\LhX}{L_{\rm HX}}
\newcommand{\Lbol}{L_{\rm bol}}

\newcommand{\kpc}{\rm kpc}

\newcommand{\ckpc}{\rm ckpc}

\newcommand{\cMpc}{\rm cMpc}

\newcommand{\lsim}{\mathrel{\hbox{\rlap{\lower.55ex\hbox{$\sim$}} \kern-.3em\raise.4ex\hbox{$<$}}}}
\newcommand{\gsim}{\mathrel{\hbox{\rlap{\lower.55ex\hbox{$\sim$}} \kern-.3em\raise.4ex\hbox{$>$}}}}

\newcommand{\Rvoid}{r_{\rm void} }
\newcommand{\REF}{Ref-L100N1504}

\newcommand{\pMpc}{\rm pMpc}
\newcommand{\eagle}{{\sc eagle}}
\newcommand{\fgas}{\rm f_{\rm \rm gas}}
\newcommand{\mhalo}{M_{\rm halo}}
\newcommand{\tlau}{\tau_{\rm 50}}

\defcitealias{schaye2015}{S15}
\defcitealias{crain2015}{C15}

\newcommand{\nii}{[N~\textsc{II}]}
\newcommand{\hii}{H~\textsc{II} }

\begin{document} 

    \title{The impact of cosmic voids on \\ AGN activity}
    \titlerunning{AGN in Voids}

   \author{Benedict L. Rouse
          \inst{1}\fnmsep\thanks{blrouse@uc.cl}
          \and
          Patricia B. Tissera\inst{1}\inst{2}
          \and 
          Yetli Rosas-Guevara\inst{3}\inst{4}
          \and
          Claudia del P. Lagos\inst{5}
          }
    \authorrunning{B.L. Rouse et al.}
   
   \institute{Institute of Astronomy, Pontificia Universidad Católica de Chile, Av. Vicuña Mackenna 4860, Santiago, 7820436, Chile.\\
              \and
             Centro de Astro-Ingeniería, Pontificia Universidad Católica de Chile, Santiago, Chile.\\
        \and 
            Donostia International Physics Centre (DIPC), Paseo Manuel de Lardizabal 4, 20018 Donostia-San Sebastian, Spain. \\
            \and
            Departamento de F\'isica, Universidad de C\'ordoba, Campus Universitario de Rabanales, Ctra. N-IV Km. 396, E-14071 C\'ordoba, Spain. \\
            \and
            International Centre for Radio Astronomy Research, The University of Western Australia, 35 Stirling Highway, Crawley, WA 6009, Australia.             }

   \date{Received ; accepted }

 
  \abstract
   {}
   {Our goal is to carry out a comparative study of the properties of central galaxies hosting active galactic nuclei (AGN) in cosmic voids and their surrounding structures (i.e. filaments and walls) at $z=0$, comparing them to non-AGN galaxies in similar environments.}
   {We used the central galaxies selected from the \eagle\ project, combined with a void catalogue that identifies voids, filaments, and walls. We categorised our sample of central galaxies into four global environments based on their distance to the nearest void. We analysed several properties such as the star formation activity and black hole mass, as a function of stellar mass and environment for galaxies with and without AGN.}
   {We found that the AGN fraction decreases as a function of void-centric distance, with void galaxies displaying the highest AGN fraction (12\%), while galaxies in denser environments, display the lowest AGN fraction (6.7\%), consistent with observations. The AGN fraction is particularly high in most massive void galaxies when controlling for stellar mass.
   When comparing AGN host galaxies to inactive ones, we find that AGN galaxies tend to have slightly more massive supermassive black holes (SMBHs), higher specific star formation rates (sSFRs), and a tendency to reside in higher mass haloes at a given stellar mass than non-AGN galaxies. At $\rm M_{*} > \rm 10^{10.2} \rm M_{\odot}$, AGN hosts in voids tend to have slightly more massive SMBHs than those in denser environments. Otherwise, the AGN population does not show a clear trend in relation to the global environment. In contrast, non-AGN void galaxies host more massive SMBHs, slightly higher sSFRs, and are located in more massive haloes than those in denser environments.
Analysing the recent merger histories of both AGN and non-AGN populations, we find that a larger fraction of massive AGN galaxies have undergone major mergers compared to non-AGN galaxies, regardless of environment. Notably, AGN galaxies in voids show a higher frequency of recent mergers (especially major mergers) than their counterparts in other environments, particularly at high stellar mass.}
   {Our results suggest that the evolution of SMBHs in voids is closely related to that of their host galaxies and their surrounding environment, while the most recent AGN activity is more strongly linked to  recent interactions.}


   \maketitle
%

\
\section{Introduction}
Supermassive black holes (SMBHs) are commonly found in the centres of nearly all massive galaxies, namely, $M_{*} \geq 10^{10} M_{\odot}$ \citep{kormendy1995,magorrian1998,ferrarese2000, gebhardt}.
They are expected to play a critical role in the regulation of the star formation activity in massive galaxies \citep[e.g.][]{bower2006,lagos2008}.
When gas is accreted onto a SMBH at large rates, it emits light as an active galactic nucleus (AGN), potentially outshining the host galaxy. Such accretion events not only grow the SMBH over cosmic time but also drive powerful winds and potentially relativistic jets \citep[e.g.][]{salpeter1964,oemler1974,dimatteo2005,hopkins2006,kormendy2013}. 

Numerical simulations of galaxy formation suggest that AGN feedback from these winds or jets could heat the surrounding gas and ultimately regulate star formation in the massive host galaxy \citep[e.g.][]{croton2006,bower2006, hopkins2006, hopkins2008, lagos2008,dubois2013, dubois2016}. Observational studies have pinpointed discrepancies among these results. Some studies argue that AGN activity is equally present in both passive galaxies and active star-forming galaxies  \citep{cardamone2010, schawinski2014}, suggesting that the AGN impact on the evolution of galaxies might be not as significant as predicted by galaxy formation models, namely, by abruptly halting star formation.  
Some studies suggest that galaxy mergers enhance AGN activity, at least for the most luminous AGN \citep{alsono2007,ellison2008A, ellison2011,treister2012,donley2018}, while others report no correlation between mergers and AGN luminosity \citep[e.g.][]{kocevski2012, villforth2014, comerford2024}. Hence, debates persist, not only as to the role of AGN feedback, but also the question of which mechanisms trigger the activity in galaxies of different masses. Incorporating the large-scale environment where galaxies are located is one interesting dimension for investigating the co-evolution of the galaxy and its SMBH.   

The large-scale structure that we observe today, the cosmic web, can be traced back to tiny density perturbations in the very early Universe \citep{lifshitz1963, weinberg1972, peebles1980}. These density perturbations have grown, mainly driven by gravity giving rise to the complex cosmic web where galaxies resides, with voids, filaments, and walls.
As the result of this evolution,  under-dense regions known as cosmic voids are also generated \citep[e.g.][]{gregory1978, einasto1980, kreckel2011}. 
Voids provide a unique laboratory to studying galaxy evolution under extreme conditions in comparison to denser regions \citep{vandeWeygaert2011}. The rate of gas accretion, interactions, and mergers are expected to be different and, hence, they affect the growth of galaxies and their SMBHs in different ways.

Observational studies from galaxy surveys have found significant differences between galaxies in void regions and those in denser environments \citep[e.g.][]{oemler1974,dressler1980,hashimoto1998, kauffmann2004, rojas2004, baldry2006, hoyle2012, kreckel2012, florez2021}. Void galaxies typically have a lower stellar mass \citep[e.g.][]{croton2005, moorman2015}, they are bluer \citep[e.g.][]{grogin2000, rojas2004, padilla2010, hoyle2012} and tend to have later type morphologies \citep[e.g.][]{rojas2004, croton2005}. However, there is still some debate ongoing over whether void galaxies differ significantly from their denser counterparts when matched for stellar mass \citep[e.g.][]{ricciardelli2014}. Studies using SDSS-DR7 have shown that void galaxies exhibit higher star formation rates (SFRs) than galaxies in void shells or control samples, although star-forming galaxies show similar activity across environments \citep{ricciardelli2014}. Consistent results were found by \citet{moorman2016} and the Void Galaxy Survey \citep{beygu2016}, where SFRs and gas content of void and non-void galaxies with similar stellar mass were comparable. However, \citet{florez2021}, using the RESOLVE and ECO surveys, found that void galaxies had more atomic gas than those in filaments and walls, suggesting a nuanced relationship between environment and gas content.
Recent results from the  Cavity Project \citep{cavityI} have shown that void galaxies exhibit a more gradual build-up compared to their counterparts in filaments, walls, and clusters, consistent with their slower star formation histories \citep{dominguezg2023}. Despite this, they appear to possess similar molecular gas fractions \citep{rodriguezcavity2024}.

In addition, the identification of cosmic voids is inherently non-trivial, as results can depend strongly on the algorithm used \citep[e.g.][]{colberg2008,neyrinck2008, cautun2018}. Comparative studies show that whilst many algorithms agree on the locations of voids, they can differ substantially in the number, sizes, and shapes of the voids they identify. 

Mergers and interactions are key physical mechanisms to drive galaxy evolution and potentially to trigger AGN feedback. Several studies have investigated the merger rate of galaxies across different environments \citep{alonso2004,deravel2011}.
Among these, \citet{perez2009} found that 50 per cent of their merging systems reside in intermediate-density environments, where a notable difference in galaxy colours has been observed \citep[see also][]{alonso2012}. \citet{ellison2010} concluded that low density environments have a higher fraction of galaxies in pairs and the highest changes in SFR, asymmetry, and bulge-total.

From a theoretical point of view, \citet{fakhouri2009} reported a strong positive correlation between merger rate and environmental density, indicating that denser environments contribute significantly to the galaxy mass build-up. \citet{lin2010} observed that mergers in dense environments have played a key role in galaxy accretion over the last 8 billion years, particularly driving the mass assembly of red galaxies. 

Conversely, \citet{kampczyk2013} found that while most galaxy pairs reside in dense regions, interactions have a stronger impact in lower-density areas, likely due to varying gas availability. Using semi-analytical models, \citet{jian2012} also confirmed a higher merger rate in denser environments, but noted that the merger timescale appears unaffected by environmental factors.

If galaxies in cosmic voids are less evolved than those in denser environments, as previous studies have suggested, then the SMBHs in void galaxies would also exhibit differences. This makes cosmic voids ideal for studying SMBHs, their activity, and their role in star formation.  However, currently there is no consensus on how the environment influences AGN activity. Some studies report that AGN activity increases with stellar mass as the environmental density increases \citep{manzer2014, argudof2018, amiri2019}, while others suggest a higher fraction of AGN-hosting galaxies in under-dense regions \citep{kauffmann2004, constantin2008, deng2012, lopes2017}. Early studies using SDSS DR2 also found that AGN are common in void galaxies \citep{constantin2008}. On the other hand, some works report no environmental dependence of  AGN fractions \citep{carter2001}. Differences in these findings may arise from the selection and classification methods for AGN. For example, some studies have shown a higher AGN fraction in dense environments \citep{manzer2014, argudof2018, amiri2019}. However, they relied on the \nii\ classification, which only distinguishes AGN, composite, and \hii regions, without identifying low-ionisation nuclear emission-line regions (LINERs), which tend to cluster more strongly than Seyferts \citep{constantin2008}. \citet{argudof2018} noted that of their 3,253 AGN, a total of 1,357 are LINER sources, which tend to disappear in bright-void galaxies, potentially skewing sample selection in under-dense regions.

Thanks to the availability of large hydrodynamical simulations \citep[e.g.][]{vogelsberger2014a, schaye2015, pillepich2018b, nelson2018}, such as \eagle\  \citep{schaye2015}, recent works have demonstrated these simulations as valuable tools for studying the large-scale environments of galaxies, given their ability to reproduce most of the observed properties of low-redshift galaxies \citep[e.g.][]{crain2015}. Many efforts have shown that \eagle\ matches well the observed AGN and SMBH properties at $z = 0$ \citep{rosasg2016}, stellar masses and SFRs \citep{furlong2015}, and gas content \citep{lagos2015,bahe2016,crain2017}.  Using \eagle, \citet{rosasg2022} found that galaxies exhibit significant differences based on their environment. For example, at high stellar mass, galaxies in denser environments are more likely to be quenched compared to those in under-dense regions. Similarly, \citet{habouzit2020}, used the \textsc{HorizonAGN} simulation, explore the properties of galaxies and BHs in voids. They found that low-stellar mass galaxies with intense star formation are more common in void interiors, with BHs and their host galaxies growing together in voids similarly to those in denser environments, despite having different growth channels.

Our study aims to carry out a comparative analysis of the properties of galaxies hosting an AGN and those without (non-AGN)  located in different large-scale environments.
For this purpose, we combine the \eagle\ hydrodynamical simulation \citep{schaye2015} and a void finder catalogue \citep{paillas2017} following  \citet{rosasg2022}.  In this paper, we analyse the differences among SMBHs residing in galaxies across various environments (i.e. voids, filaments, and walls). The focus will be on disentangling the effect of the environment on the triggering of AGN activity and the AGN feedback effects on the host galaxies by distinguishing the properties of galaxies hosting AGN from those without and investigating whether void properties influence AGN activity and key properties of their hosts.

This paper is organised as follows. Section \ref{sec:data} describes the simulation and the classification of galaxies according to environment. Section \ref{sec:hostgalprops} studies the properties of the AGN and their host galaxies at $z = 0$. Section \ref{sec:propsovertime} describes the evolution of AGN 
properties over cosmic time. The main results are summarised in Section \ref{sec:conclusions}.

\section{Methodology}

\label{sec:data}
\subsection{Simulations}
\label{sec:sims}

We used the \eagle~project\footnote{Details about the \eagle~project can be found at the project's website:  \\ \citet[][\url{http://eagle.strw.leidenuniv.nl}]{mcalpine2016}} \citep{schaye2015, crain2015}, which comprises a suite of cosmological simulations performed with different subgrid physics and different numerical resolutions and volumes.
These simulations were executed using a modified version of P-Gadget-3, an upgraded version of Gadget-2 \citep{springel2005}. This modified version integrates changes to the vanilla smoothed particle hydrodynamics, referred to as {\sc ANARCHY}, enabling the incorporation of unresolved physical phenomena such as cooling, metal enrichment, star formation activity, and energy input from stellar sources \citep{schaye2008}, as well as black hole growth \citep{rosasg2015}. For a detailed look at the comparison of hydrodynamics schemes, we refer to \citet{schaller2015}. The simulations are consistent with a Lambda cold dark matter ($\Lambda$-CDM) cosmology with the following cosmological parameters: dark energy density of  $\Omega_\Lambda=0.693$, matter density of $\Omega_{\rm m}=0.307$, baryon density of $\Omega_{\rm b}=0.04825$, matter fluctuations of $\sigma_8=0.8288$, Hubble normalisation of $h=0.6777$, scalar spectrum power-law index of $n_{s}=0.9611,$ and a fraction of baryonic mass in helium $Y=0.248$ \citep{planck2014}.

In particular, we analyse the  largest volume denoted as \REF~also studied by \citet{rosasg2022}.
The \REF~encompasses a comoving volume of $(100\, \cMpc)^3$. The simulation has a mass resolution of $9.7 \times 10^6 \,\Msun$ for dark matter (DM) particles and $1.81 \times 10^6 \,\Msun$ for baryonic particles, with a corresponding comoving softening length of $2.66 \,\ckpc$, limited to a maximum physical size of $0.70\, \kpc$.

\subsection{AGN in \eagle}
\label{subsec:agncriterion}
The inclusion of BHs in the \eagle\ simulations is fully described in \citet{schaye2015} and \citet{rosasg2016}. Hence, we only highlight the main relevant information for our work below. A galaxy is seeded with a BH of mass $1.48 \times 10^5\Msun$, if the halo of the host galaxy exceeds a mass of $1.48 \times 10^{10} \Msun$. This BH may grow through gas accretion. $\dot{M}_{\rm acc}$ is the mass accretion rate onto a SMBH, which is calculated by the modified Bondi-Hoyle model \citep{rosasg2015}:
\begin{equation}
    \dot{M}_{\rm acc} = min(\dot{M}_{\rm bondi}[C^{-1}_{\rm visc}(c_{\rm s}/V_{\Phi})^{3}], \dot{M}_{\rm bondi}),
\end{equation} 
where $C^{-1}_{\rm visc}$ is a dimensionless free parameter, $c_{\rm s}$ the sound speed, and $V_{\Phi}$, the circular velocity of the surrounding gas.
\begin{table}
\caption{ Characteristics of the analysed samples.}
    \centering
    \scalebox{0.9}{\begin{tabular}{c|c|c|c}
    \hline
         Environment & AGN count & Galaxy count & AGN percentage (\%) \\
         \hline
         In-V & 61 & 492 & 12.4 $\pm$ 1.5\\
         Out-V & 47 & 460 & 10.2 $\pm$ 1.4\\
         W & 261 & 3082 & 8.5 $\pm$ 0.5\\
         S & 81 & 1208 & 6.7 $\pm$ 0.7\\
         \hline
    \end{tabular}}
    \tablefoot{From left to right columns: the environment, the total number of central AGN galaxies, the total number of central galaxies and the percentage of these galaxies that host an AGN in the corresponding environment in the \REF~\eagle\ simulation, at $z = 0$. Errors are from the binomial confidence intervals, calculated as $
        \sigma = \sqrt{\frac{p(1-p)}{n}}$, where p is the fraction of AGN and n is the galaxy count.}
    
    \label{tab:numbers}
\end{table}
In our work, we used the Eddington ratio and X-ray luminosity to define where a galaxy hosts an AGN. The Eddington ratio is defined by \begin{equation}
    \Ledd \equiv \rm log_{10}(\dot{\textit{M}}_{\rm accr}/\dot{\textit{M}}_{\rm Edd}), \label{eq:edd}
\end{equation}
where the Eddington accretion rate, $\dot{M}_{\rm Edd}$ is defined as 
\begin{equation}
    \dot{M}_{\rm Edd} = 4\pi G M_{\rm BH} m_{\rm p} / \epsilon_{\rm r} c \sigma_{T}, 
\end{equation}
where $G$ represents the gravitational constant, $c$ the speed of light in vacuum, $m_{\rm p}$ the mass of the proton, and $\sigma_T$ the Thompson cross-section. Also, $\epsilon_{\rm r}$ represents the radiative accretion efficiency. $M_{\rm BH}$ is the mass of the BH.     
We identify a galaxy as hosting an AGN if $\rm log_{10}(\Ledd)\geq -3$ and the (hard) X-ray luminosity is $L_{\rm X}$ $\geq$ $10^{40}$ erg$s^{-1}$.
To calculate the hard X-ray luminosity, we must first calculate the bolometric luminosity ($\Lbol$) as
\begin{equation}
    \Lbol = \frac{\epsilon_{r}}{1 - \epsilon_{r}} \dot{M}_{\rm BH} c^{2} = \epsilon_{r} \dot{M}_{\rm acc} c^{2},
\end{equation}
with the accretion efficiency, $\epsilon_{\rm r} = 0.1$, inline with the work of \citet{shakura1973}. Then, the correction for intrinsic hard X-ray luminosity (2-10 keV) from \citet{marconi2004}. The conversion from $\Lbol$ to $\LhX$ is done via the third degree polynomial,
\begin{equation}
    \rm {log_{10}}\left(\frac{\textit{L}_{HX}}{\textit{L}_{bol}}\right) = -1.54 - 0.24\textit{L} - 0.012\textit{L}^{2} + 0.0015\textit{L}^{3},
\end{equation}
requiring \(\Ledd \geq 10^{-3}\) enables us to identify more AGN than selecting only those with radiatively efficient accretion flows (\(\Ledd \geq 10^{-2}\)) or super-Eddington accretion (\(\Ledd \geq 1\)), while still ensuring higher accretion rates than those associated with advection-dominated accretion flows (ADAFs; \(\Ledd \sim 10^{-4}\)). This criterion allows us to achieve more robust statistical analyses across the four defined environments. The X-ray luminosities of our sample span from $10^{40}$ erg $\rm s^{-1}$ to $10^{44}$ ergs $\rm s^{-1}$.

\subsection{The selected \eagle\ galaxy sample}
\label{sec:params_def}
The galaxies were identified utilising SUBFIND to detect gravitationally bound sub-structures \citep{springel2001,dolag2009} within each  friends-of-friends (FoF) halo. In this study, we only analysed central galaxies defined as the most massive sub-structure within the DM halo, which were part of the sample analysed by \citet{rosasg2022}.
The stellar masses are quantified within a fixed aperture of 30 physical  $\kpc$. 
The galaxies that remain within the FoF halo are classified as satellite galaxies and will not be considered in this analysis. In the subsequent sections, we analyse the following galaxy parameters: 
\begin{itemize}
    \item {$\mbh$}: the sum of all BH particles in the halo;
    \item {$\mhalo$}: the total mass contained within the virial radius ($R_{200}$), within which the average density of the halo is 200 times the critical density of the Universe; 
    \item sSFR:  the ratio between the SFR, and the stellar mass within a 30 pkpc aperture. The SFR is estimated by using the star-forming gas in the \eagle\ simulation, and thus corresponds to an instantaneous measurement of the SFR;
    \item $\fgas$: the gas fraction is defined as the ratio between the total gas mass in a physical 30 $\kpc$ aperture and the stellar mass of a galaxy;
    \item $\tlau (M_{*})$:  the look-back time at which 50\% of the z = 0 galaxy stellar mass  was assembled;
    \item $\tlau (M_{\rm halo})$: the look-back time at which 50\% of the z = 0 halo mass was assembled. 
    
\end{itemize}
We used the $\mbh$ values provided in the public database, defined as the total mass of all BHs in the halo. \citet{mcalpine2016} and \citet{lagos2025} demonstrated that when $\mbh \geq 10^{6} \rm M_{\odot}$, this value is approximately equal to the most massive BH. Thus, when we refer to BH masses throughout this work, we are referring to the total value as opposed to the central.

\subsection{Voids in \eagle}
\label{subsec:voidsineagle}
We worked with the void catalogue constructed by \cite{paillas2017} \footnote{\citet{paillas2017} discussed the impact of varying the free parameters. The  public catalogue was built with their best combination. }
 where galaxies with a stellar mass of $\rm M_* \geq 10^8\Msun$ are used as void tracers.
The catalogue also encompasses filaments and walls structures.
In brief, voids are found with a spherical under-density finder, based on the algorithm presented in \cite{padilla2005}. The algorithm identifies spherical under-densities in a galaxy distribution by building a rectangular space grid and counting the number of galaxies in each cell of the grid. The empty cell centres in the grid are candidates to be void centres. Spherical under-density profiles are calculated around each candidate up to the integrated density exceeding 20 per cent of the mean galaxy number density. The radius of the largest sphere used in the calculation is defined as the void radius, $\Rvoid$. If two void candidates have similar centres and volumes and their centres are closer than 40 per cent of the sum of their radii, the smallest of them is discarded. To verify $\Rvoid$ for the remaining voids, the void centre is shifted in different directions, and if the new radius is larger than $\Rvoid$, the position of the void is updated. The void catalogue consists of 709 voids, with nine of  having  $\Rvoid$ values greater than $10\,\pMpc$. The largest void is $\sim 24\,\pMpc$ and the smallest void is $\sim 5\,\pMpc$.

We assigned a simulated galaxy to one of four specific environments depending on its distance from the centre of the void, using the same definitions as in \citet[][see their Fig.1]{rosasg2022} as follows:
\begin{enumerate}
    \item inner void, In-V: if a galaxy is located between the centre of the void and $\Rvoid = 0.8$, it is defined as an inner void galaxy; 
    \item outer void, Out-V: if the galaxy lies between $0.8\,\Rvoid$ and $\Rvoid$, it is in the outer void region; 
    \item wall, W: for galaxies located between $\Rvoid$ and $1.4\,\Rvoid$, they are labelled as wall galaxies; 
    \item skeleton, S: any galaxy located beyond $1.4\,\Rvoid$ from the centre of the void is classified as a skeleton galaxy.
\end{enumerate}

An important feature of this void catalogue is the size of the voids with respect to SDSS measurements and catalogues from larger simulations.   \citet{paillas2017} reported that the size  of the largest voids in the \eagle~simulations are smaller than $\approx 25$ Mpc, which is in global agreement with observations. In addition, these authors found  that their largest void in their sample is comparable to the smallest void in the simulation box used by \citet{hamaus2014} and \citet{cai2015}. Hence, we cannot probe very large voids in this study. The definition of voids and the tracers used to identify them and how this affects the 
statistics have been studied in detail in \citet{paillas2017}.

\subsection{AGN samples in different environments}
\label{subsec:agnineagle}

\begin{figure}

  \begin{subfigure}[b]{1\columnwidth}

        \includegraphics[width=1.01\textwidth, trim = 15 0 40 10, clip]{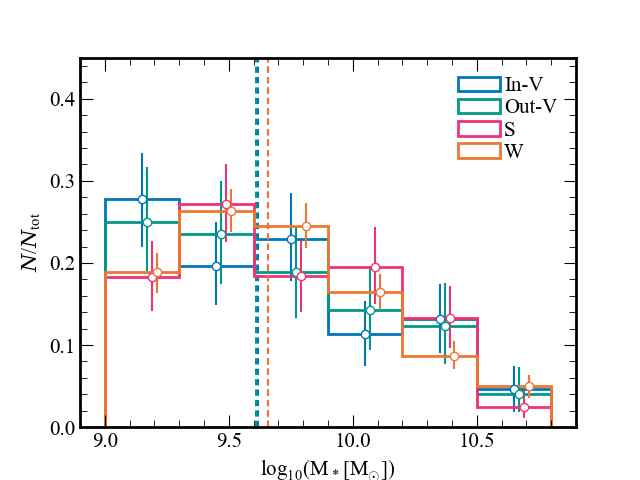}
  
  \end{subfigure}
  \begin{subfigure}[b]{1\columnwidth}

        \includegraphics[width=1.01\textwidth, trim = 15 0 40 10, clip]{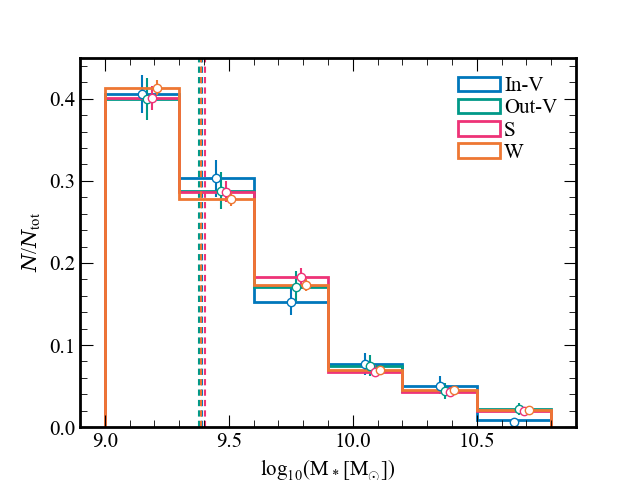}

  \end{subfigure}
  \caption{Stellar mass distribution of the AGN  (top panel) and the non-AGN galaxies (bottom panel) in the inner void (blue),  outer void (green),  skeleton (magenta), and wall (orange) cosmic environments. The vertical lines denote the median stellar mass for each environment. We note that the AGN and non-AGN samples have been drawn from galaxy samples in each environment that have matching stellar mass distributions \citep{rosasg2022}. Error bars show the 16th and 84th percentiles from the bootstrap distribution of 1000 samples.}
  \label{fig:mass_dist}
\end{figure}

Using the criteria defined in Section~\ref{subsec:agncriterion}, we selected AGN from the sample of galaxies in each environment defined  by \citet{rosasg2022},  which were constrained to have the same stellar mass distribution in each of the four defined environments. 
The numbers of AGN and non-AGN galaxies, as well as the percentage of galaxies hosting an AGN in each environment are summarised in Table~\ref{tab:numbers}. The AGN percentage exhibits a noticeable increase in under-dense environments, ranging from $12.4$ per cent for inner void galaxies to $6.7$ per cent for skeleton galaxies. Furthermore, we present the mass distribution of both the AGN and non-AGN at $z=0$ in Fig.~\ref{fig:mass_dist}.  Across all environments, AGN consistently exhibit a larger median stellar mass compared to non-AGN (see vertical lines in the top versus bottom panel of Fig.~\ref{fig:mass_dist}). This is a similar result to that found in \citet{dubois2015}, where they reported that  AGN in low-mass galaxies were suppressed due to supernova feedback  \citep[see also][]{bower2017}. Furthermore, in Fig.~\ref{fig:mass_dist_halo} we show that the distribution of halo mass is quite comparable to the stellar mass distributions. AGN are found to reside in larger haloes, regardless of the environment than galaxies without an AGN at $z=0$.

\begin{figure}  
   
    \includegraphics[width=1.05\columnwidth, trim = 15 0 20 10, clip]{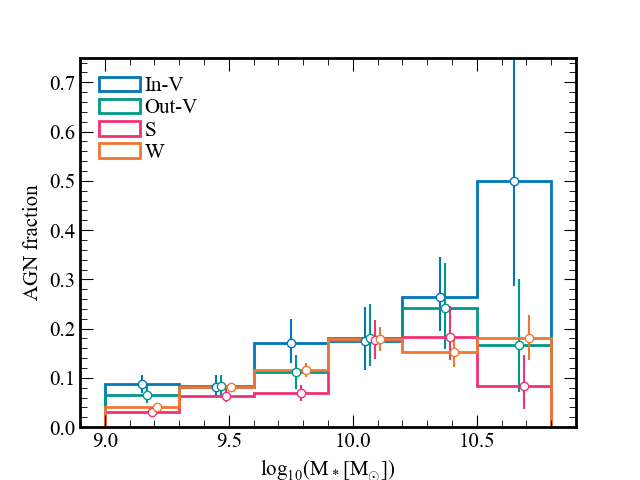}
    \caption{AGN fraction as a function of stellar mass for galaxies in  each defined environment. 
    Galaxies in under-dense environments are more likely to host an AGN than those in denser environments,  particularly for  $\rm M_{*}  \sim 10^{10.5} \rm M_{\odot}$, at $z=0$. Errors show the 16th and 84th percentiles from the bootstrap distribution of 1000 samples.}
    \label{fig:AGNfraction}
\end{figure}

 We show the AGN fraction per stellar mass bin per environment in Fig.~\ref{fig:AGNfraction}. 
 From this figure, it can be seen  that the AGN fraction is consistently higher in the In-V regions compared to other environments across various stellar mass bins. Furthermore, in the highest stellar mass bin ($\rm log_{10}(\rm M_{*}[\rm M_{\odot}]) > 10.5$), around half of the galaxies in the In-V regions are hosting an AGN at $z = 0$. This suggests that there is a connection between environment and AGN activity. 
 These results align with those presented by \cite{constantin2008}, who reported a higher rate of AGN occurrence in void regions compared to walls, particularly in massive galaxies, through the analysis of spectroscopic and photometric data from the SDSS DR2 catalogue. Likewise, \cite{habouzit2020} utilised the Horizon-AGN simulation and found a greater AGN fraction in void galaxies compared to shell galaxies.

In the subsequent sections, we delve into the characteristics of galaxies harbouring AGN and those devoid of AGN activity. Our aim is to understand the elevated AGN activity detected in void regions and identify the potential mechanisms behind this phenomenon.

\section{AGN and their host galaxy properties at $z=0$}
 
\label{sec:hostgalprops}

This section will specifically examine the characteristics of the AGN host galaxy, comparing them across the four environments defined in Section \ref{subsec:voidsineagle} and with non-AGN galaxies as well.  In particular, we focus on the galaxy properties defined in Section \ref{sec:sims}. To carry out this analysis, the galaxies in each environment were separated into the following stellar mass intervals: 
\begin{itemize}
    \item {low mass} : 9 $\leq$ $\rm log_{10}(\rm M_{*} [\rm M_{\odot}]) <$ 9.5; 
    \item {intermediate mass} : 9.5 $\leq$ $\rm log_{10}(\rm M_{*} [\rm M_{\odot}]) <$ 10; 
    \item {high mass} : $\rm log_{10}(\rm M_{*} [\rm M_{\odot}]) \geq 10.$
\end{itemize}

\begin{figure}
    \hspace{-0.41cm}
    \includegraphics[width=1.12\columnwidth, trim = 15 0 20 10, clip]{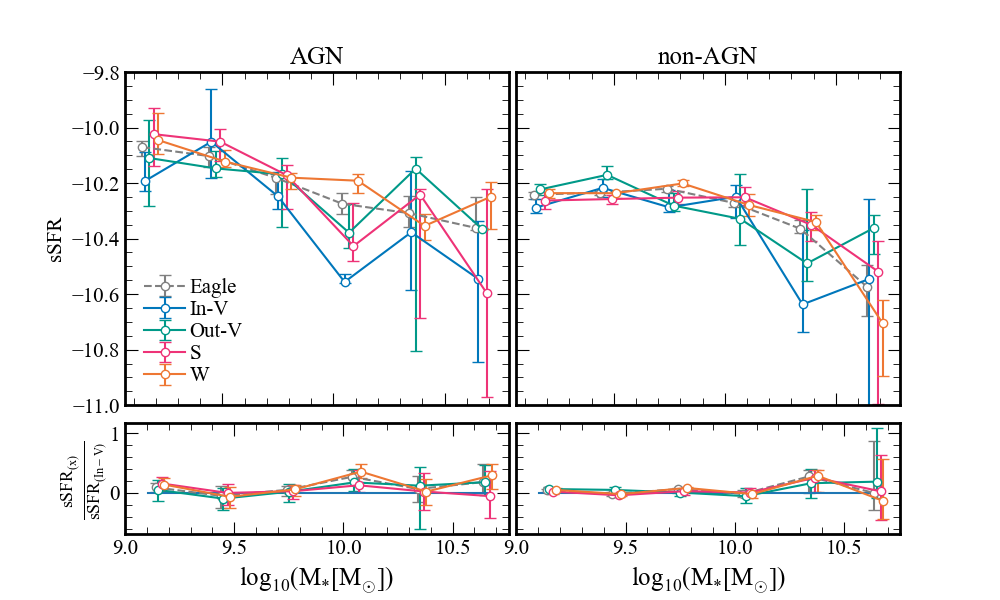}
    \caption{\textit{Top}:  Log sSFR as a function of stellar mass for AGN (left) and non-AGN (right). The circle points are the median value of the log sSFR in that stellar mass range. The error bars denote the bootstrap errors in each stellar mass interval. Each point has been slightly shifted along the x-axis so that the error bars are visible. \textit{Bottom left}:  Ratio between the log sSFR of Out-V, S, W, and all AGN host galaxies to In-V. \textit{Bottom right}: Ratio between the log sSFR of Out-V, S, W, and all non-AGN host galaxies to In-V (see Section \ref{subsec:voidsineagle} for definitions). }
    \label{fig:comp_SFR_boot}
\end{figure}

\begin{figure}
    
    \hspace{-0.22cm}
    \includegraphics[width=1.1\columnwidth, trim = 15 0 20 10, clip]
    {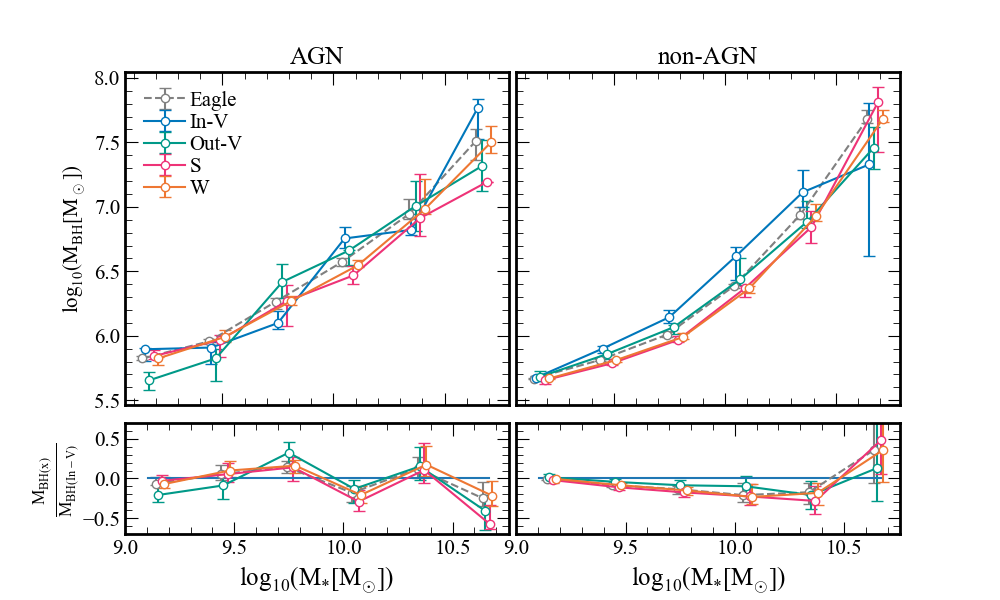}
    \caption{Same as Fig. 
    \ref{fig:comp_SFR_boot} for log $\mbh$ as a function of stellar mass.} 
    \label{fig:comp_bhmass}
\end{figure}

\begin{figure}
    \hspace{-0.22cm}
    \includegraphics[width=1.1\columnwidth, trim = 15 0 20 10, clip]{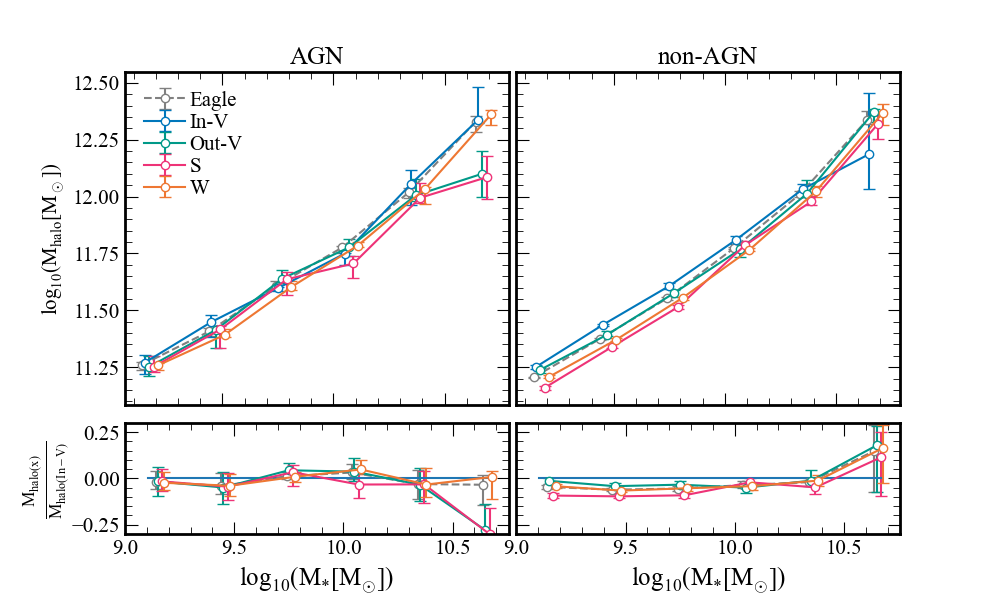}
    \caption{Same as Fig. 
    \ref{fig:comp_SFR_boot} for the log($\mstar$)-log($\mhalo$) relation.}
    \label{fig:comp_halomass}
\end{figure}

\begin{figure}
    \hspace{-0.22cm}
    \includegraphics[width=1.1\columnwidth, trim = 15 0 20 10, clip]{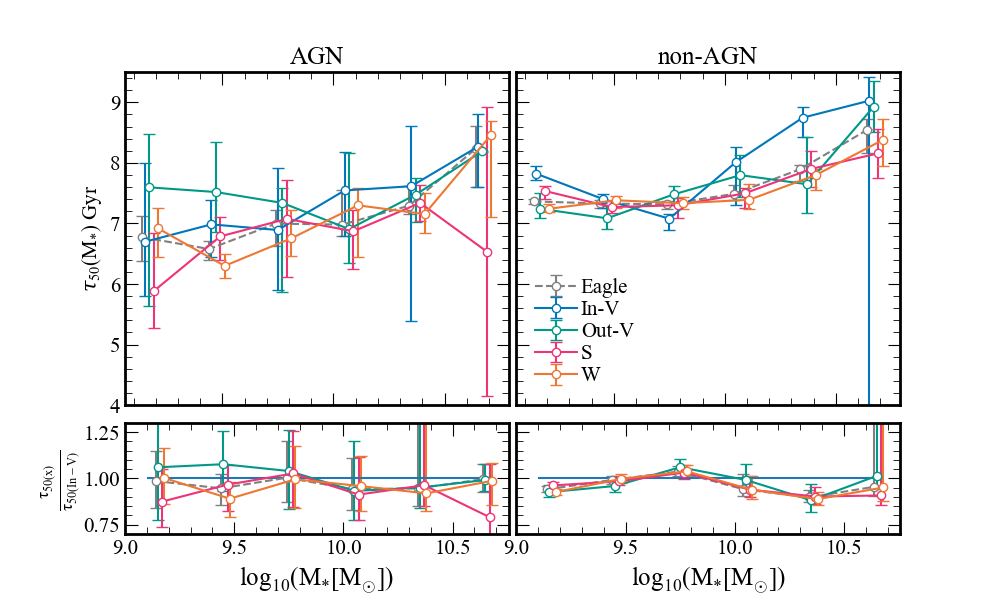}
    \caption{Same as Fig. 
    \ref{fig:comp_SFR_boot} for the look-back time at which 50 per cent of the stellar mass was formed as a function of stellar mass.} 
    \label{fig:comp_lbt50}
\end{figure}

In Figs. \ref{fig:comp_SFR_boot}, \ref{fig:comp_bhmass}, \ref{fig:comp_halomass}, and  \ref{fig:comp_lbt50}, we  give the main parameters of AGN (left panel) and non-AGN (right panel) galaxies as a function of their stellar mass in the four defined environments (coloured-coded lines), overlaid with the parameter distribution of the entire sample for AGN and non-AGN, respectively (grey dashed lines). The bottom panels show the medians of the corresponding parameter divided by the medians of the In-V AGN (bottom left) and non-AGN (bottom right) as a function of stellar mass. Hence, we  define $\frac{x_{env}}{x_{In-V}}$, where $x$ is the galaxy parameter and $env$ is another environment (not excluding the entire sample in grey).

Figure~\ref{fig:comp_SFR_boot} depicts the sSFR as a function of $\mstar$ for both AGN and non-AGN in the four different environments. 
AGN systems seem to determine a stronger sSFR-$M_{*}$ anti-correlation for  masses in the range 9.0 $< \rm log_{10}(M_{*}[M_{\odot}]) <$ 10.0 than galaxies that do not host an AGN. 
For stellar masses higher than $10^{10} \rm M_{\odot}$, AGN galaxies seems to reach a plateau. However, we warn about the large dispersion, as indicated by the error bars. 

The sSFR in In-V galaxies tends to be slightly lower than in the other environments, across the entire mass range and in  particular, for AGN galaxies, as can be seen in both of the bottom panels of Fig.~\ref{fig:comp_SFR_boot} and Table~\ref{tab:galparams}. 
From this table, it is also clear that low mass AGN systems have systematically higher sSFR than their non-AGN counterparts for all masses and in all environments. 

Combining all environments, our results show that AGN in \eagle~prefer host galaxies with high sSFR, in low mass and the highest mass bin, as can be seen with the grey lines in both plots, in general agreement with the findings of \citet{ward2022}. These authors looked across three different  simulations: IllustrisTNG \citep{springel2018}, \eagle\  \citep{schaye2015}, and SIMBA \citep{dave2019}. They found that AGN preferentially reside in galaxies with high sSFRs and a high molecular gas fraction (see their Fig. 1).

Figure~\ref{fig:comp_bhmass} displays total $\mbh$ versus $\mstar$ for the AGN and non-AGN populations across the four environments. We appreciate that overall, AGN have slightly more massive $\mbh$ values at a given stellar mass than non-AGN (see the second column of Table~\ref{tab:galparams}). This trend holds at stellar masses below $10^{10.5}M_\odot$; however, at the highest stellar masses it appears to reverse, particularly in the lowest-density environments.

As can be shown in the bottom left panel, the $\mbh$ of AGN shows significant scatter with environment across the whole stellar mass range, while non-AGN in In-V have systematically more massive BH. The exception might be the most massive non-AGN galaxies, where there is a reversal in the trend. However, since it is only one mass bin and the error bars are large, we cannot draw any strong conclusion.

For $\mstar > 10^{9.5}\rm \Msun$, it seems that AGN have been able to feed their $\mbh$ more efficiently than non-AGN, as can be seen from the difference in the median $\mbh$ in Table~\ref{tab:galparams}.
From this table and Fig. \ref{fig:comp_bhmass} (upper panels), it can be seen  that, regardless of stellar mass, AGN in W and S environments have $\mbh$ values about $0.1-0.3$ dex higher than non-AGN in the same environments. In contrast, for AGN and non-AGN in under-dense regions, the difference in $\mbh$ is much smaller. Therefore, although In-V galaxies have a higher likelihood of hosting an
AGN, they appear to have been less efficient at feeding their SMBHs in comparison
to those in a higher density environment, except for the higher mass bin,
In-V galaxies have the same AGN-to-non-AGN $\mbh$ ratio. 

Upon examining the median $\mbh$ values shown in Table~\ref{tab:galparams}, it becomes evident that there is lower variation in $\mbh$ of AGN and non-AGN in low-mass galaxies. The current masses closely resemble the initial seeding mass in the \eagle\ simulation ($\approx 10^{5}\rm M_{\odot}$). This implies that these small galaxies have experienced minimal growth and, by construction of the subgrid model, they are also less affected by AGN feedback. We notice that this corresponds to the stellar mass range where supernovae suppress BH growth in  \eagle~ \citep{bower2017}. Therefore, any conclusions drawn from this stellar mass interval should be taken with caution.

In Fig.~\ref{fig:comp_halomass}, we show the $\mhalo-\mstar$ relation. At a given stellar mass, galaxies hosting an AGN tend to reside in slightly more massive dark matter haloes than non-AGN galaxies, a trend that is also reflected in Table~\ref{tab:galparams}. 
This can also be appreciated in Fig. \ref{fig:mass_dist_halo}, which shows the halo mass distribution in AGN (top panel) and non-AGN (bottom panel) across different environments. 

Since halo mass is closely tied to the assembly history of galaxies -- with more massive haloes generally forming earlier and experiencing different accretion histories -- this could suggest that AGN host galaxies have undergone a distinct evolutionary pathway. In particular, the presence of an AGN in more massive haloes could point to earlier or more sustained periods of gas accretion and feedback, potentially leading to a different regulation of star formation over cosmic time. Thus, the observed halo mass difference could be a consequence of, or a contributor to, the divergent star formation histories of AGN and non-AGN galaxies. \citep[e.g.][]{tremonti2007,silk2012, fabian2012,bower2017,beckmann2017, weinberger2018, dave2019}. 

\begin{figure*}
  
  \begin{subfigure}[b]{0.3\textwidth}
    
        \includegraphics[width=1.3\columnwidth]{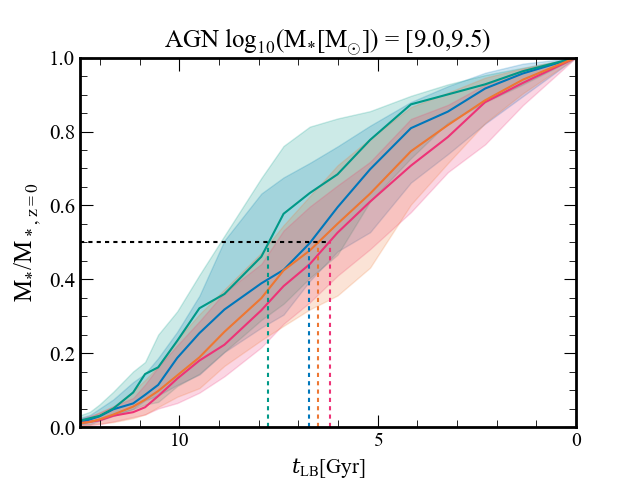}
  
  \end{subfigure}
  \begin{subfigure}[b]{0.3\textwidth}
    
        \includegraphics[width=1.3\columnwidth]{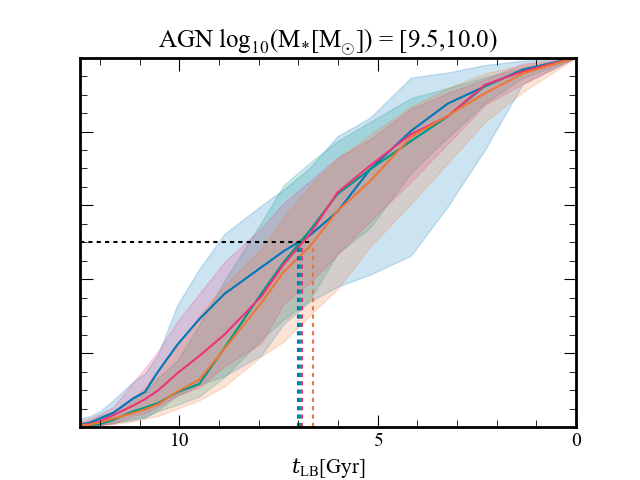}

  \end{subfigure}
  \begin{subfigure}[b]{0.3\textwidth}
    
        \includegraphics[width=1.3\columnwidth]{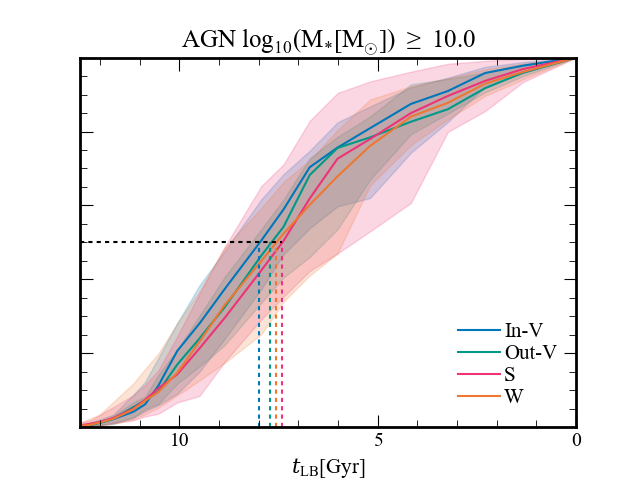}

  \end{subfigure}

  \begin{subfigure}[b]{0.3\textwidth}

        \includegraphics[width=1.3\columnwidth]{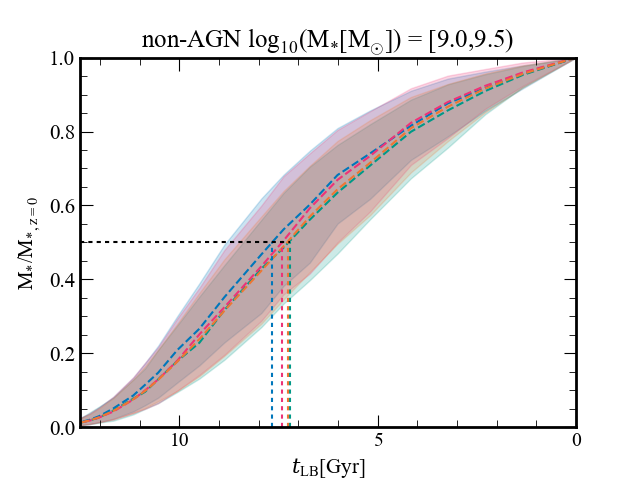}

  \end{subfigure}
  \begin{subfigure}[b]{0.3\textwidth}
    
        \includegraphics[width=1.3\columnwidth]{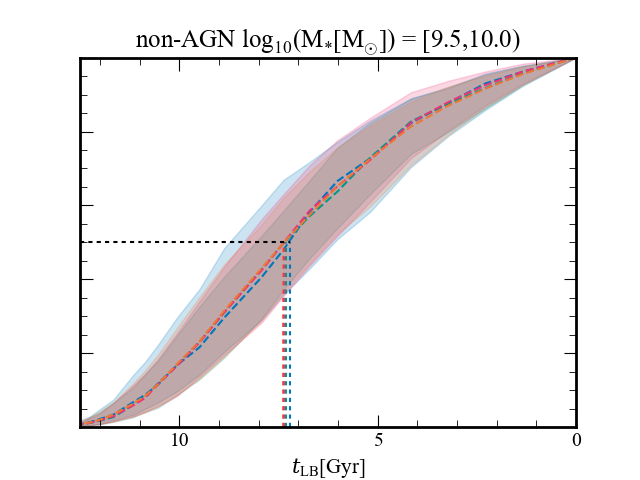}

  \end{subfigure}
  \begin{subfigure}[b]{0.3\textwidth}
    
        \includegraphics[width=1.3\columnwidth]{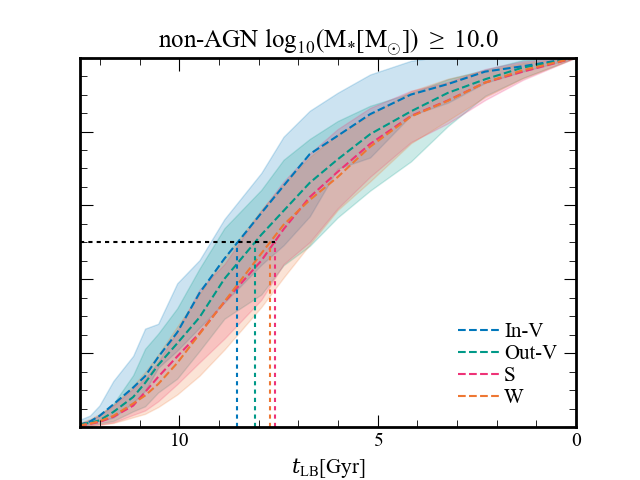}

  \end{subfigure}
  
  \caption{Stellar mass at a given $z$ relative to the final stellar mass at $z=0 $ for AGN (upper panels) and non-AGN (lower panels) galaxies for galaxies in different stellar mass ranges. The mean formation time of the galaxies, estimated as the look-back time at which  galaxies formed 50 per cent of their final stellar mass, $\tlau$ are included for each subsample (dashed lines).  AGN and non-AGN central galaxies formed quicker in lower In-V and Out-V regions over W and S, particularly in the low and high stellar mass intervals. AGN central galaxies are assembled systematically later than non-AGN ones.}

  \label{fig:stellarmassbuildup_AGN}
\end{figure*}
We go on to focus on non-AGN systems and compare their halo mass in different environments at a fixed stellar mass, as shown in the bottom panel of Fig.~\ref{fig:comp_halomass}.  We notice that a trend for galaxies in the In-V environment to reside in higher mass haloes than their counterparts in Out-V, W, or S environments (i.e. the halo mass slightly decreases with increasing environmental density). In other words, galaxies at a given $\mhalo$ host more massive galaxies in denser regions in agreement with previous works \citep{artale2018, rosasg2022}. However, this trend changes for galaxies hosting AGN.
As can be seen in the middle panel of Fig.~\ref{fig:comp_halomass}, 
AGN in different environments have similar DM halo masses, except for very massive systems where there is a large dispersion for both AGN and non-AGN.

These results suggest that AGN activity seems to reduce the difference in halo mass among environments, while non-AGN show systematically higher $\mhalo$ and $\mbh$ for galaxies in voids than non-AGN in Out-V, W, and S, in agreement with the results of \citet{rosasg2022}. The similarity between  the trends shown in the bottom two panels of both Figs. \ref{fig:comp_bhmass} and \ref{fig:comp_halomass} and Table~\ref{tab:galparams} suggest that AGN reside in slightly more massive haloes and this allows for higher accretion rates onto the SMBH regardless of the global environment. This is in agreement with the findings of \cite{bower2017} who discussed the translation of the growth of SMBHs into a relation between $\mbh$ and $\mhalo$, providing insights into the interplay between BH growth and halo growth in a cosmological context in the \eagle\ simulations. Analytical models have been proposed to explain the origin of the characteristic mass scale in galaxy formation, which is observed as an abrupt change in galaxy properties around a halo mass of $10^{12}\Msun$ \citep{dubois2015,habouzit2017, angles2017, hopkins2022}. The model allows for a connection between the halo growth rate, the effectiveness of star formation-driven outflows, and the gas density in the galaxy. The density of gas surrounding the SMBH is shown to depend critically on the effectiveness of feedback from star formation. The model is supported by observational data and confirmed in numerical experiments.

Finally, in Fig.~\ref{fig:comp_lbt50} we display $\tlau(M_{*})$ as a function of the stellar mass.  It is clear that the AGN population has a larger variation in $\tlau$, indicating a wider diversity of formation histories in comparison with non-AGN galaxies.  We found a tendency for AGN-hosting galaxies to exhibit lower $\tlau$ values of about $0.5$ Gyr, on average. 
This implies that galaxies hosting AGN typically have more recent stellar formation and/or accretion histories than those without AGN, as can also be seen in  Table~\ref{tab:galparams}. 

\section{AGN properties over cosmic time}
\label{sec:propsovertime}
In this section, we analyse the cosmic evolution of key properties of AGN and non-AGN galaxy populations. We also describe their recent merger events in the different environments.

\subsection{Assembly histories for AGN and non-AGN populations}

\begin{figure*}

  \begin{subfigure}[b]{0.3\textwidth}
    
        \includegraphics[width=1.3\columnwidth]{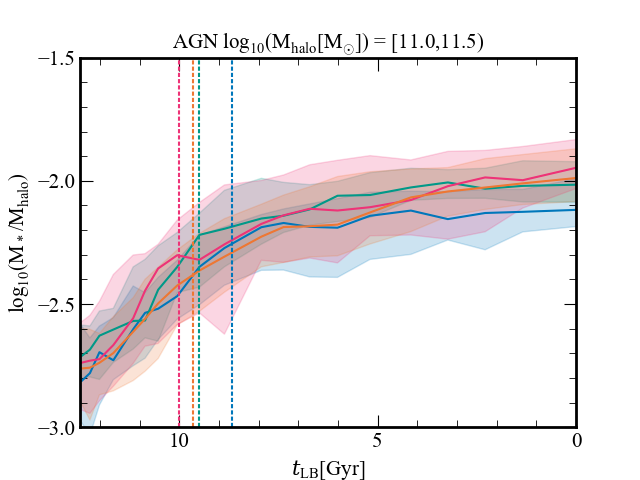}
  
  \end{subfigure}
  \begin{subfigure}[b]{0.3\textwidth}

        \includegraphics[width=1.3\columnwidth]{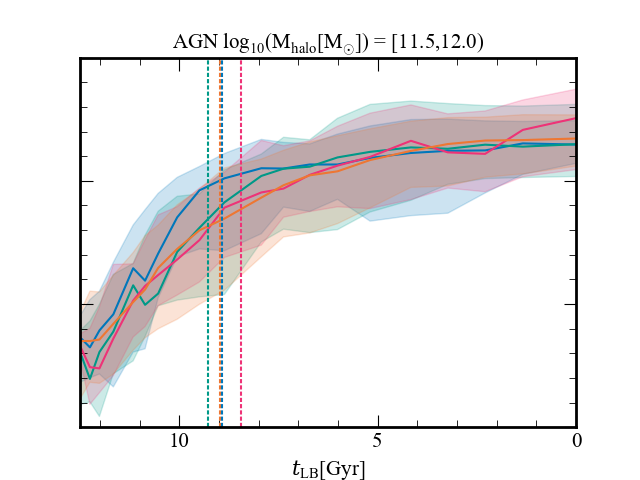}

  \end{subfigure}
  \begin{subfigure}[b]{0.3\textwidth}

        \includegraphics[width=1.3\columnwidth]{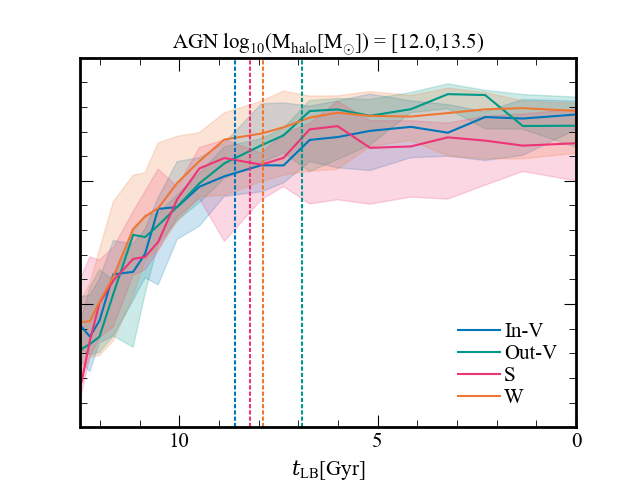}

  \end{subfigure}
  
  \begin{subfigure}[b]{0.3\textwidth}
    
        \includegraphics[width=1.3\columnwidth]{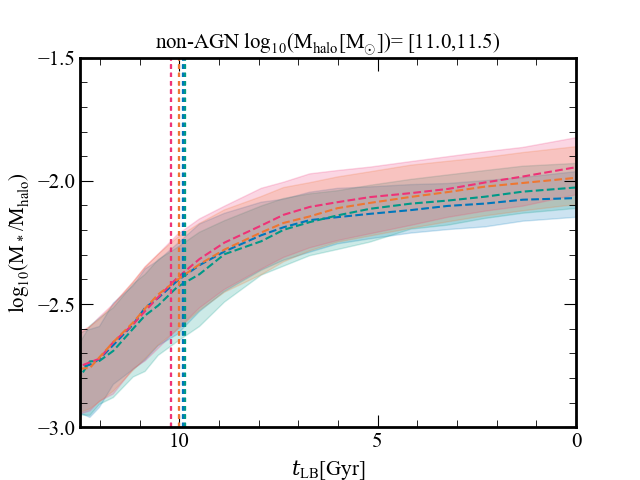}
  
  \end{subfigure}
  \begin{subfigure}[b]{0.3\textwidth}
    
        \includegraphics[width=1.3\columnwidth]{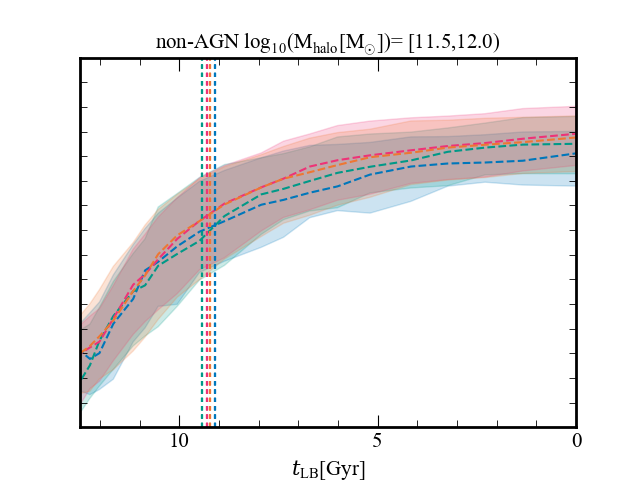}

  \end{subfigure}
  \begin{subfigure}[b]{0.3\textwidth}
    
        \includegraphics[width=1.3\columnwidth]{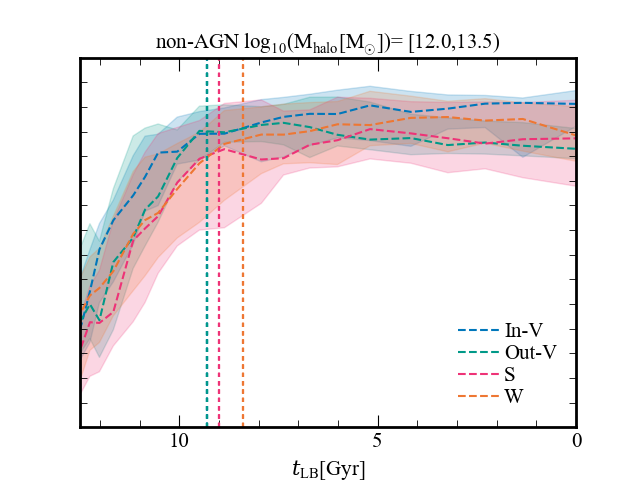}

  \end{subfigure}
  \caption{Evolution of $\mstar$-$\mhalo$ ratio over look-back time for  AGN (top panel) and non-AGN  (bottom panel) host galaxies. The look-back times at which the haloes reached $50$ per cent of its final halo mass are included (dashed lines). From left to right: Increasing $\mhalo$ bins.}
  \label{fig:stellarhalobuildup}
\end{figure*}

We estimated the mass assembly histories by following the main progenitor of the $z=0$ galaxies back in time.
In the upper panel of Fig.~\ref{fig:stellarmassbuildup_AGN} we display the stellar mass build-up of galaxies that host an AGN for three stellar mass intervals, each split into the four defined environments.  The dotted vertical lines depict $\tlau (M_{*})$ given in Table~\ref{tab:galparams}. As can be seen,  AGN galaxies in In-V and Out-V assembled $50$ per cent of their final stellar mass before those in walls and skeletons. This is also in agreement with the results of Fig.~\ref{fig:comp_lbt50}.

Galaxies hosting an AGN in under-dense regions in the highest mass interval also seem to grow faster than those in walls or skeletons until they reach 50 per cent of their current stellar mass, and then their growth is similar across different environments. 
Because the stellar mass distribution are similar for AGN hosts in the  analysed environments with only a slight higher fraction of low mass galaxies in under-dense regions with low stellar mass, this trend suggests that at a given stellar mass, the assembly was slightly faster in under-dense regions at high redshift.

We also display the stellar mass assembly of the non-AGN galaxy population in the lower panel of Fig.~\ref{fig:stellarmassbuildup_AGN}. With the exception of the most massive galaxies, we observe smaller variations in the mass growth of non-AGN galaxies across different environments. This figure reinforces the observation that non-AGN galaxies consistently exhibit higher $\tlau$ values compared to AGN galaxies in the same environment, implying earlier assembly histories, as it was already mentioned.
As shown earlier in Fig.~\ref{fig:comp_lbt50} and Table~\ref{tab:galparams}, in each stellar mass bin, galaxies lacking an AGN assembled 50 per cent of their stellar mass earlier than those with an AGN. AGN host galaxies are also forming stars with a higher efficiency at $z=0$. These differences could be explained by the action of AGN \citep{rosito2019} possibly as well as also supernovae feedback, which  regulate (and possibly delay) the transformation of gas into  stars in the progenitors.

 \citet{rodriguez2024} reported that void galaxies are younger than galaxies in other environments using the TNG300 simulation, which might be in tension with our findings in \eagle.
However, we should take into account that these authors considered both central and satellite galaxies, whereas we only included central galaxies. The total mass accumulation was undistinguishable in all environments when they solely considered central galaxies. Furthermore, TNG300 has a higher number of large void regions than our sample in \eagle. However, the TNG300 numerical resolution is lower than \eagle. Similarly, through observations, \cite{dominguezg2023} discovered that when compared to galaxies in other environments, void galaxies formed stars slower compared to cluster environments. However, these authors did not distinguish between central and satellite galaxies or AGN and non-AGN. In a follow-up study, \citet{torres2024} separated their sample into central and satellite galaxies and find that both local and large-scale environments play a relevant role in shaping the star formation history of galaxies. They also report that group galaxies, especially those in voids and with lower stellar mass, tend to assemble their stellar mass earlier than isolated (singlet) galaxies. 
 \begin{figure*}
  \includegraphics[width=1.07\linewidth, trim = 30 400 50 260, clip]{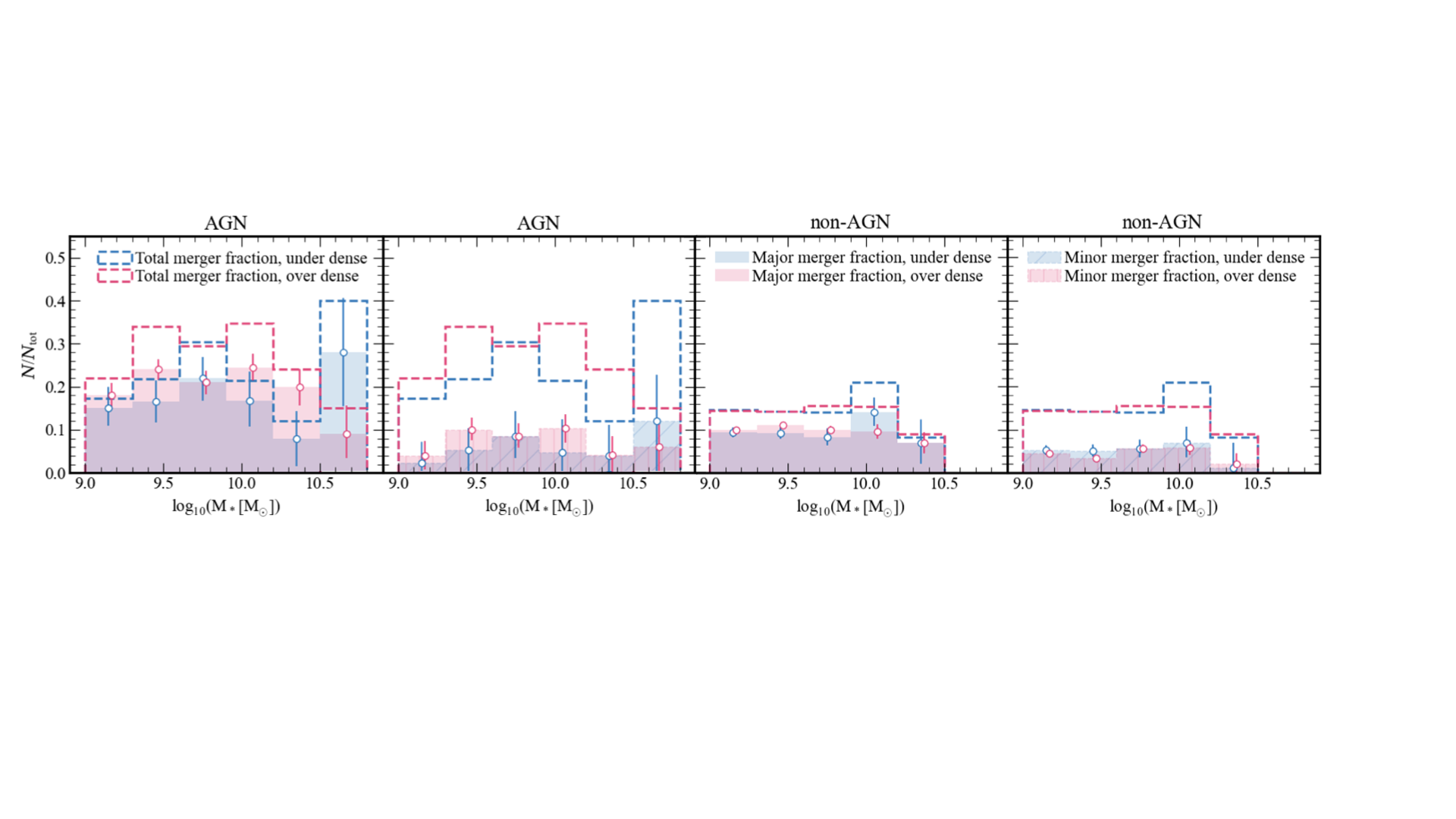}
  \caption{Total (dashed lines), major (shaded areas), and minor (hatched shading) merger fraction as a function of $\mstar$ for AGN (left two panels) and non-AGN (right two panels) host galaxies in under-dense (blue lines and shades) and over-dense (magenta lines and shades) environments at $z = 0$. The solid shaded regions in the left panels represent major mergers, while the dashed lines indicate the total merger fraction within each stellar mass bin. The hatched shaded regions in the right panels correspond to the minor mergers. Error bars show the 16th and 84th percentiles from the bootstrap distribution of 1000 samples. We also tested jackknife errors and found that the bootstrap errors were larger.}
  \label{fig:merger}
\end{figure*}
We also analyse the $\fgas$ (defined in Sect. 
\ref{sec:params_def}), finding  very little difference in the $\fgas$ for AGN galaxies regardless of the large-scale environment (see Sect.~\ref{sec:gas_appendix}). The $\fgas$ evolution is also very similar across all environments for galaxies that do not host. However, AGN have systematically higher sSFR at $z =0 $, which implies that they are currently  more efficiently at transforming gas into stars and fuelling it to feed their BHs than non-AGN. These trends agree with recent observational results from CO-Cavity survey by \citet{rodriguezcavity2024}.

Another crucial aspect to consider is the stellar mass-to-halo mass ratio, which we show in Fig. \ref{fig:stellarhalobuildup}. This can be interpreted as a proxy for galaxy formation efficiency per dark matter halo.  
As reported above, at $z=0$ and at a given stellar mass, non-AGN galaxies of a given stellar mass  range are found to inhabit slightly less massive haloes than AGN in a given environment. To investigate the time evolution of the $\mstar/\mhalo$  ratio in Fig.~\ref{fig:stellarhalobuildup} we show the $\mstar/\mhalo$ as a function of the look-back time for various halo mass bins. 

As can be seen, the evolution of $\mstar/\mhalo$  for all samples show a significant variation regardless of environment. To get a better quantification of the efficiency, we calculated the time at which 50 per cent of the DM halo is assembled, $\tau_{50}$($M_{\rm halo}$). These times are depicted in Fig.~\ref{fig:stellarhalobuildup} by the vertical dashed lines, coloured by their respective environments.

These estimations show a slight trend for DM haloes to assemble latter if they host an AGN and no clear dependence on the environment, except for low mass galaxies.
There is no significant difference between DM haloes on environment for a given stellar mass range, although we found that AGN reside in slightly higher DM halo masses. 

\subsection{The role of recent mergers}

In this section, we explore the impact of mergers in relation to the activation of the AGN. Mergers are well-known mechanisms that can induce  gas inflows to the central regions, triggering star formation activity \citep{barnes1996, tissera2000} and possible feeding the SMBHs \citep{hopkins2006}. Hence, we analysed the relation with the most recent violent event. We adopted the classification of last minor and/or last major merger parameters used by \citet{rosasg2022}. They measured the stellar mass ratio of the merging pairs, $\mu$, by determining the stellar mass of both the merging galaxy and the primary galaxy in the last snapshot before the merger event. If the merging galaxy had a value in the range of $\mu =[0.1,0.25],$ it was defined as a minor merger. In cases where $\mu \geq$ 0.25, it was defined as a major merger. When discussing mergers and merger fractions, it is important to clarify that we did not consider the number of mergers each galaxy has experienced across its formation path. Instead, we focussed solely on the look-back time of its most recent merger, applying an upper limit of $\rm t < 2.4 \rm\ Gyr$ to explore the potential link between the merger event and the relevance of the AGN activity.  

 To increase the statistics, in this section, galaxies are categorised into two broad environments,  under-dense (i.e. inner and outer voids together) and over-dense (i.e. wall and skeleton regions together) instead of the original four ones. This reorganisation of galaxies enables us to conduct more robust statistical analyses when galaxies are also selected by the time since the last merger event.

In Fig. \ref{fig:merger}, we present the fraction of galaxies that have recently undergone a merger. 
The left two panels show the merger fractions for AGN host galaxies, while the merger fractions for non-AGN  galaxies are shown in the bottom panels. 
Major mergers are depicted as shaded bars, while in the right panels, minor mergers are shown as shaded and hatched bars. The total merger fraction is denoted by the dashed line. Recent mergers and especially major mergers seem to be more frequent in AGN galaxies than in their non-AGN counterparts,  by approximately a factor of two, in agreement with what has been reported in previous observational works \citep[e.g.][]{ellison2011,ellison2013,gao2020,comerford2024}. Non-AGN galaxies show a larger fraction of minor mergers. This suggests that at least in \eagle\ galaxies, major mergers might be more efficient in triggering AGN activity. 

We want to draw attention to the highest mass bin in AGN galaxies, which exhibits a significant spike in the merger fraction among under-dense galaxies, while  galaxies in over-dense regions show a decrease in mergers. A similar  spike in AGN fraction in inner voids was reported in Fig.~\ref{fig:AGNfraction}. This suggests that potentially (and at least for the more massive galaxies major mergers in under-dense regions) could be an  efficient mechanism for switching on an AGN, at least for the more massive galaxies. 

Observationally, \citet{comerford2024} noticed that the AGN excess occurs more frequently in high-mass galaxies that have experienced a recent merger. They also note that the AGN excess is slightly stronger in major mergers than in minor mergers. \citet{perez2025}, using the morphological information of Galaxy Zoo 2 galaxies \citep{domsan2018}, found that minor mergers are less frequent in low density environments; however, their study was based on voids with larger minimum and smaller maximum radii than those in our catalogue. Finally, using TNG300, \citet{rodriguez2024} found that void galaxies tend to experience mergers at later times. Figure~\ref{fig:merger} shows this trend specifically for galaxies hosting an AGN. However, for non-AGN galaxies, we did not observe significant differences between under-dense and over-dense environments, within the uncertainties in the \eagle\ simulations.


\section{Conclusions}
\label{sec:conclusions}
Using the largest $\Lambda$CDM cosmological hydrodynamical simulation from the \eagle\ project \citep{schaye2015,crain2015}, we examined the properties of AGN and non-AGN in relation to their large-scale environment. We employed the void catalogues of \citet{paillas2017}. Using the void-centric distance, we used central galaxy samples with comparable stellar mass distributions in the four defined environments \citep{rosasg2022}. 

We categorised the galaxies according to their AGN activity using the Eddington ratio (Eq. \ref{eq:edd}) and X-ray luminosity. We note that cluster environments are not included in our analysis. 
Although it is possible to study the effect of the large-scale environment in the evolution of galaxies and their SMBH, there are several caveats. Given the simulated volume of the \eagle\ simulation, our sample does not include voids with sizes larger than 25pMpc, whereas observed voids could extend to larger radii \citep{hoyle2012, ricciardelli2014}. Additionally, the poor time cadence of snapshots of \eagle\ makes it is difficult to follow the AGN activity in time \citep{rosas2019},  probably missing  active events. However, when comparing our results with the general galaxy properties in different environments, we find similar trends. Our main results can be summarised as follows:

\begin{itemize}
     \item We find that the fraction of AGN decreases as a function of increasing void centric distance. Inner void galaxies exhibit the highest AGN fraction (12\%) and galaxies in the denser skeleton environment present the lowest AGN fraction (6.7\%), as shown in Table \ref{tab:numbers}. The AGN fraction of void galaxies is the highest in the most massive galaxies (Fig. \ref{fig:AGNfraction}). This is consistent with observations \citep[e.g.][]{ceccarelli2022} and other cosmological simulations \citep[][]{habouzit2020}.

 \item Regardless of the large-scale environment in which they reside, galaxies that host an AGN generally exhibit a higher sSFR than non-AGN galaxies at $z=0$ (see Table \ref{tab:galparams}). For galaxies with  $M_{*}[M_{\odot}]<10^{9.5}$, there is a minor tendency for void galaxies to have a higher sSFR. Although the dispersion in the sSFR-stellar mass diagram is high for all environments (Fig \ref{fig:comp_SFR_boot}).

 \item We find  that galaxies with AGN have more massive BHs at a given stellar mass than non-AGN galaxies. Nevertheless, AGN galaxies that reside in voids and have a $M_{*}>10^{9.5}M_{\odot}$  exhibit marginally higher massive BHs than the other environments (Fig. \ref{fig:comp_bhmass}) and AGN in voids with $M_{*}>10^{10.2}M_{\odot}$ have the largest mass difference with non-AGN in voids. We also notice that non-AGN galaxies in voids have more massive BHs than those in denser environments across the entire stellar mass range, excluding those with ($M_{*}>10^{10.5}M_{\odot}$).   

\item Galaxies with AGN typically inhabit more massive haloes than those without AGN at a given stellar mass. However, we did find a significant dependence of AGN halo mass on environment. Conversely, non-AGN galaxies tend to have systematically less massive haloes in denser environments (Fig. \ref{fig:comp_halomass}).

     \item AGN host galaxies tend to assembly their stellar mass later in time than their non-AGN counterparts.  There is a slight tendency for galaxies to be  assembled earlier in voids rather than in other denser environments. This seems to contradict the observations (Fig. \ref{fig:stellarmassbuildup_AGN}). However, we note that we estimate the assembly histories by following the progenitor systems back in time, while observations  reconstruct a history by using stellar populations. 
     Our results agree with those of \citet{rodriguez2024}, who discovered no distinction in galaxies when they exclusively examined central galaxies in the TNG300 simulation. A prospective analysis that includes satellite galaxies in the \eagle\ simulations should be conducted to evaluate the distinctions. Additionally, our \eagle\ sample does not include cluster environments.

\item  When we explored the evolution and the halo mass and halo-stellar mass relation, we found that  non-AGN galaxies in voids tend to reside  in more massive haloes than those in dense environments since early times (Fig. \ref{fig:stellarhalobuildup}). 
However, AGN show almost no dependence of the $\mstar-\mhalo$ value on environment and have slightly more massive haloes than non-AGN systems at a given $\mstar$. 
In addition, haloes in voids assembled slightly earlier than haloes in other environments, suggesting that local environment is more critical than large-scale environment for recent AGN activity. 

\item  We investigated the fraction of galaxies that have recently ($t < 2.4 \rm Gyr$) undergone a major or minor merger by defining two global environments: high and low density. We found that the fraction of galaxies that have undergone a recent major merger is higher for AGN compared to non-AGN regardless of environment (Fig. \ref{fig:merger}).  This suggests that the recent AGN activity of the galaxies may be ascribed to the recent mergers, at least in part, rather than a cause of the large-scale structure in which AGN reside. However, there is a particularly large spike in merger activity in the highest mass, low-density environment AGN host galaxies, in line with the higher AGN fraction in high mass In-V galaxies (Fig. \ref{fig:AGNfraction}), potentially shining light onto the most efficient method of feeding an AGN, at low redshift.

\end{itemize}

With the advent of upcoming and ongoing cosmological surveys, such as the deep-field $\it{Roman}$ telescope (\citet{akeson2019} and $\it{Euclid}$, with a lower resolution, but wider field of view \citep{laureijs2011,euclid2024}, it will become possible to place further constraints on galaxy evolution models in relation to void environments. Coupled with AGN surveys such as $\it{e-Rosita}$ \citep{predehl2021} and synergies with other surveys such as J-PAS \citep{benitez2014} and Cavity \citep{perez2024}, these efforts will provide deeper insights into the growth and evolution of SMBHs. Additionally, new high-resolution, large-volume cosmological simulations will enable a better understanding of these connections and allow for more accurate comparisons with galaxies residing in observed voids.

\begin{acknowledgements}
      PBT acknowledges partial funding by Fondecyt-ANID 1240465/2024, and ANID Basal Project FB210003.  This project has received funding from the European Union Horizon 2020 Research and Innovation Programme under the Marie Sklodowska-Curie grant agreement No 734374- LACEGAL. 
We acknowledge the use of  the Ladgerda Cluster (Fondecyt 1200703/2020). 
\end{acknowledgements}

%
\bibliographystyle{aa} 
\bibliography{biblio.bib} 
%

\begin{appendix}
\section{Halo mass distribution}
\begin{figure}
  \begin{subfigure}[b]{1\columnwidth}
    
        \includegraphics[width=1\textwidth]{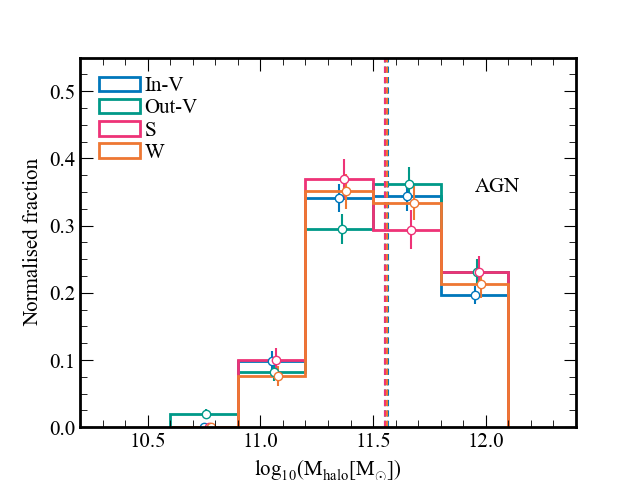}
  
  \end{subfigure}
  \begin{subfigure}[b]{1\columnwidth}
    
        \includegraphics[width=1\textwidth]{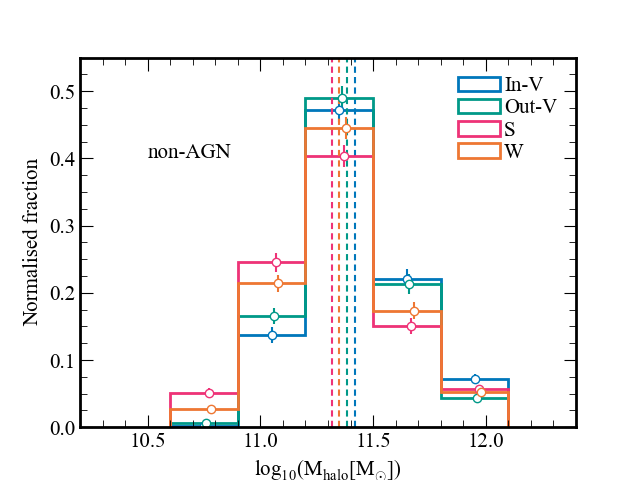}

  \end{subfigure}
  \caption{Normalised histograms of the halo mass distribution of AGN  (top panel) and non-AGN galaxies (bottom panel) in the inner void (blue),  outer void (green),  skeleton (magenta), and wall (orange) cosmic environments, at $z=0$. The dashed vertical lines indicate the median halo mass, for each environment. We note that the AGN and non-AGN samples have been drawn from galaxy samples in each environment that had matching stellar mass distributions \citep{rosasg2022}.}
  \label{fig:mass_dist_halo}
\end{figure}

In Fig. \ref{fig:mass_dist_halo}, we plot the halo mass distributions of AGN (top) and non-AGN (bottom). Similarly to Fig.\ref{fig:mass_dist}, where we plot the stellar mass distributions for the two samples, we notice that the AGN host galaxies are more likely to reside in heavier haloes than their non-AGN counterparts. For AGN, there is very little dependence on environment, whereas the non-AGN in under-dense environments tend to populate the higher end of the mass distribution than the over-dense non-AGN galaxies. In \citet{fontanot2011}, they find that for central galaxies, similar to the galaxies in our sample, AGN reside in more massive galaxies and haloes than star-forming galaxies (SFG); whereas, when only satellite galaxies are considered, the mass distribution of both stellar and halo and much more similar for AGN and SFGs. This could be due to the inefficient quenching in galaxies that reside in massive haloes, providing a rich reservoir of material for the central BH to accrete and produce an AGN. 

\section{Gas fractions}
\label{sec:gas_appendix}
A key ingredient for the star formation and AGN activity is the availability of gas. We analysed the gas fraction, $\fgas$, as a function of look-back time finding that the evolution of the gas fraction for the AGN host galaxies is similar, regardless of the large-scale environment, except for the galaxies in the lowest stellar mass bin.  For low-mass AGN  galaxies in the inner and outer void regions, the median gas fraction peaks $10$ Gyr ago and sharply decreases today. Whereas, there is a lower peak in the AGN hosts in dense environments, which seems to have remained about flat until today (see Fig. \ref{fig:gasfracbuildup}). 
\begin{figure}
    
    \hspace{-0.41cm}
    \includegraphics[width=1.12\columnwidth, trim = 15 0 20 10, clip]{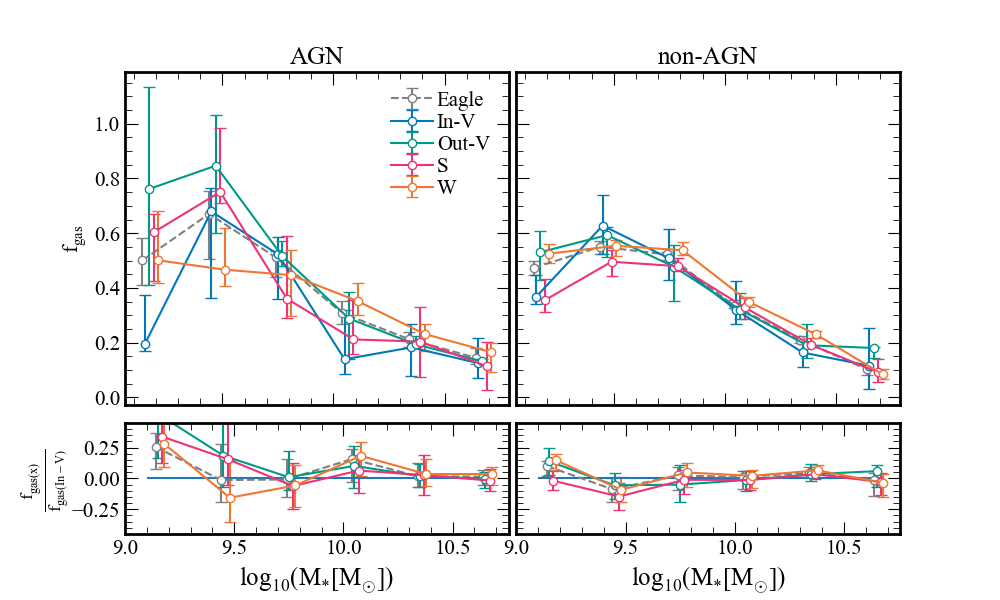}
    \caption{Same as Fig. \ref{fig:comp_SFR_boot} for Gas fraction, ($f_{\rm gas} = \frac{\rm M_{\rm gas}}{\rm M_{*}}$), as a function of stellar mass for AGN (\textit{left}) and non-AGN (\textit{right}). }
    \label{fig:comp_gasfrac}
\end{figure}

\begin{figure*}
  
  \begin{subfigure}[b]{0.3\textwidth}
    
        \includegraphics[width=1.3\columnwidth]{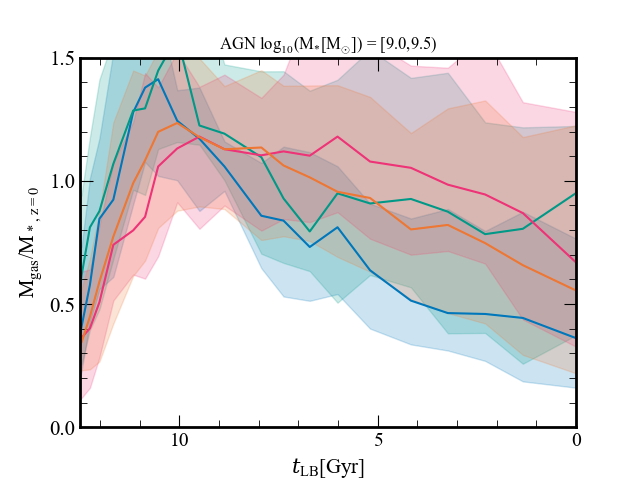}
  
  \end{subfigure}
  \begin{subfigure}[b]{0.3\textwidth}

        \includegraphics[width=1.3\columnwidth]{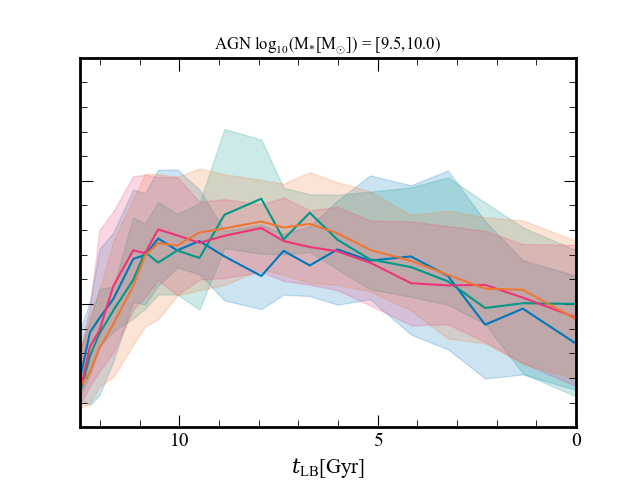}

  \end{subfigure}
  \begin{subfigure}[b]{0.3\textwidth}

        \includegraphics[width=1.3\columnwidth]{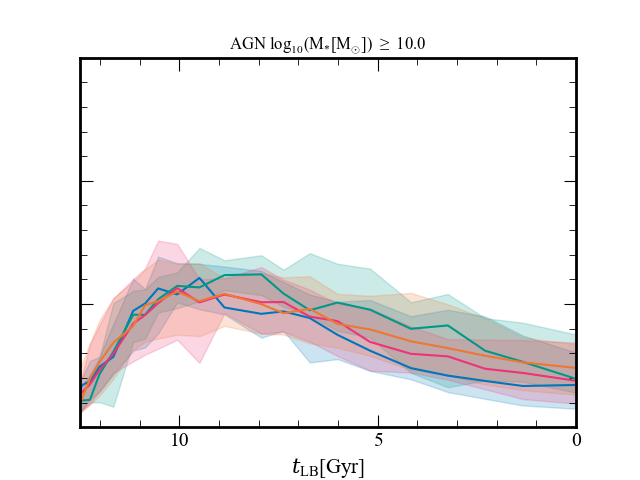}

  \end{subfigure}
  
  \begin{subfigure}[b]{0.3\textwidth}
    
        \includegraphics[width=1.3\columnwidth]{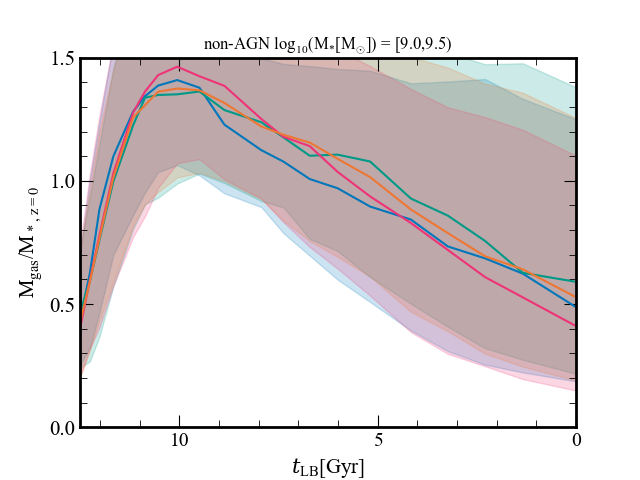}
  
  \end{subfigure}
  \begin{subfigure}[b]{0.3\textwidth}
    
        \includegraphics[width=1.3\columnwidth]{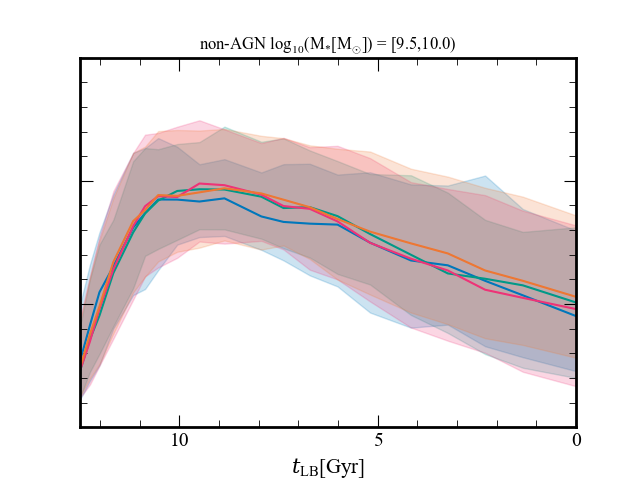}

  \end{subfigure}
  \begin{subfigure}[b]{0.3\textwidth}
    
        \includegraphics[width=1.3\columnwidth]{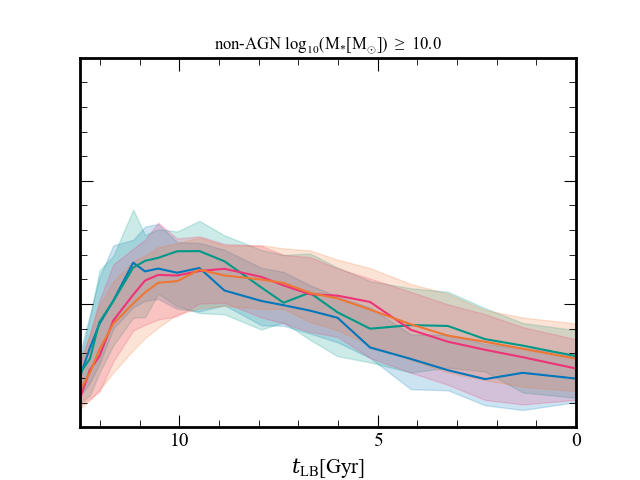}

  \end{subfigure}
  \caption{Top panel: AGN. Bottom: Non-AGN. Left to right: increasing stellar mass bins. The evolution of the gas mass over the stellar mass of the galaxy at z=0 as a function of look-back time. In the two higher mass bins, there is very little difference between the environment or between the AGN and non-AGN in the build-up of gas mass. In the smallest stellar mass bin, we need to carry out some statistical tests or something similar to understand the differences.}
  \label{fig:gasfracbuildup}
\end{figure*}

\begin{table}[]
    \centering
    \begin{tabular}{c|c|c}
       Env  & AGN & Non-AGN \\
       \midrule
        In-V  & ${ 0.67 ^{+ 0.61 }_{- 0.34 }}$  & ${ 0.41 ^{+ 0.69 }_{- 0.26 }}$   \\[0.1cm]
    Out-V   &${ 0.95 ^{+ 0.27 }_{- 0.58 }}$  & ${ 0.59 ^{+ 0.79 }_{- 0.37 }}$   \\[0.1cm]
    
    W  &${ 0.55 ^{+ 0.67 }_{- 0.34 }}$  & ${ 0.53 ^{+ 0.73 }_{- 0.34 }}$  \\[0.1cm]
    
    S  &${ 0.67 ^{+ 0.61 }_{- 0.34 }}$  & ${ 0.41 ^{+ 0.69 }_{- 0.26 }}$  \\[0.1cm]
    \bottomrule

    In-V  &${ 0.34 ^{+ 0.28 }_{- 0.19 }}$  & ${ 0.45 ^{+ 0.35 }_{- 0.22 }}$ \\[0.1cm]
    Out-V  &${ 0.50 ^{+ 0.22 }_{- 0.38 }}$  & ${ 0.51 ^{+ 0.31 }_{- 0.31 }}$ \\[0.1cm]
    W  &${ 0.44 ^{+ 0.32 }_{- 0.24 }}$  & ${ 0.53 ^{+ 0.33 }_{- 0.25 }}$   \\[0.1cm]
    S  &${ 0.45 ^{+ 0.30 }_{- 0.28 }}$  & ${ 0.48 ^{+ 0.34 }_{- 0.31 }}$  \\[0.1cm]
    \bottomrule

    In-V  &${ 0.17 ^{+ 0.13 }_{- 0.10 }}$  & ${ 0.20 ^{+ 0.10 }_{- 0.10 }}$  \\[0.1cm]

    Out-V &${ 0.20 ^{+ 0.18 }_{- 0.06 }}$  & ${ 0.29 ^{+ 0.11 }_{- 0.17 }}$     \\[0.1cm]
    
    W  &${ 0.24 ^{+ 0.10 }_{- 0.11 }}$  & ${ 0.28 ^{+ 0.14 }_{- 0.13 }}$  \\[0.1cm]
    
    S  &${ 0.19 ^{+ 0.15 }_{- 0.09 }}$  & ${ 0.24 ^{+ 0.12 }_{- 0.13 }}$   \\[0.1cm] 
    \end{tabular}
    \caption{Same as Table \ref{tab:galparams}, but solely for the $\fgas$ parameter. Split into three different stellar mass intervals:  9.0 < log($M_{*}$) < 9.5 (top panel),  9.5 < log($M_{*}$) < 10.0 (middle panel), and  log($M_{*}$) > 10.0 (bottom panel). }
    \label{tab:gas_frac}
\end{table}
The evolution of the gas fraction for non-AGN galaxies is shown in the bottom panel of Fig.~\ref{fig:gasfracbuildup}. Although the drop in the gas fraction is much more consistent in inactive galaxies, the evolution is similar in all the environments. It is also worth noting that the scatter in the evolution is considerable, indicating a high degree of variability in the evolution of the gas fraction. This is not entirely surprising, as the gas content of galaxies is a very dynamical property \citep{crain2017}.

In Fig. \ref{fig:comp_gasfrac} we look at the fraction of gas $f_{\rm gas}$ in the galaxy. Here, we notice that In-V galaxies which host an AGN, at low stellar mass, tend to have a slightly lower $f_{\rm gas}$ than the other environments. However, when compared to non-AGN, there seems to be very little difference between the $f_{\rm gas}$ in galaxies that host an AGN, and galaxies that do not. This is an initial indicator that the amount of gas that a galaxy possesses is not the factor driving our AGN fractions in Fig \ref{fig:AGNfraction}.
The larger errors present in the AGN sample arise from there being a lower sample size. As can be seen from the bottom panel, however, the median values for both AGN and non-AGN galaxies are very similar across the stellar mass range, despite the large errors. \citet{ellison2019} find that in a sample of SDSS AGN,  when the samples are fixed within stellar mass and SFR ranges, in high mass galaxies ($\rm log_{10}(\rm M_{*}/\rm M_{\odot}) > 10.2$) the AGN sample has a similar HI fraction to inactive galaxies. However, in the mass range of $9.6 < \rm log(\rm M_{*}/\rm M_{\odot}) < 10.2$ the AGN are HI poor.  In their lowest mass sample, $9.0 < \rm log(\rm M_{*}/\rm M_{\odot}) < 9.6$, the AGN host HI content has a deficit by a factor of 4. In all but the highest mass range, our results here differ. This could be due to the different gas and stellar mass measurements required in simulations and observational work. In \citet{ward2022}, they look within three different cosmological simulations to understand the gas fraction and sSFR in galaxies that host an AGN and those that do not. They find that AGN are preferentially found in galaxies with high $f_{\rm gas}$ and sSFR. Also, the fraction of AGN that are depleted in gas and have quenched star formation is lower than in galaxies where there is no AGN. Clearly, we agree that AGN are preferentially found in galaxies with higher sSFR, but disagree that they are found in high $f_{\rm gas}$ galaxies. Table \ref{tab:gas_frac} shows that there is only an increase in $f_{\rm gas}$ in the low-mass AGN host galaxies, but must be deemed inconclusive due to the errors.

\section{Median properties of AGN and non-AGN galaxies}
Here, we provide a table of median properties of AGN and non-AGN host galaxies to support the figures presented in Sect.~\ref{sec:hostgalprops}.
\begin{sidewaystable*}
\caption{Median properties of AGN and non-AGN galaxies in different environments}

  \scalebox{0.96}{%

    \begin{tabular}{lSSSSSSSSSSSSSSSS}
    
    \toprule
    \multirow{2}{*}{Env.} &
      \multicolumn{4}{c}{$\text{sSFR}$ ($10^{-11} \rm s^{-1}$)}  &
      \multicolumn{4}{c}{$\rm log_{10}$($\mbh$) [$\rm M_{\odot}$]} &
      \multicolumn{4}{c}{$\rm log_{10}$($\mhalo$) [$\rm M_{\odot}$]} &
      \multicolumn{4}{c}{$\tlau$ ($M_{*}$) [Gyr]}\\
      & {AGN} &  & {non-AGN} & & {AGN} & & {non-AGN} & & {AGN} & & {non-AGN}& &{AGN} & &{non-AGN}& \\
    \midrule
      In-V & ${ 8.16 ^{+ 3.45 }_{- 1.22}}$ &{1.12}  & ${ 5.66 ^{+ 2.41 }_{- 3.55}}$  & {0.34}  & ${ 5.87 ^{+ 0.47 }_{-0.53 }}$ & {0.06} & ${ 5.72 ^{+ 0.45 }_{-0.28 }}$ & {0.02} &  ${ 11.33 ^{+ 0.77 }_{-0.60 }}$&{0.03}  & ${ 11.30 ^{+ 0.71 }_{-0.55 }}$&{0.01}  & ${ 6.70 ^{+ 1.24 }_{- 2.18 }}$ &{0.58}  & ${ 7.69 ^{+ 1.28 }_{- 1.13 }}$ &{0.12}\\[0.1cm]
    Out-V & ${ 6.85 ^{+ 2.17}_{- 5.21 }}$&{1.81}  & ${ 6.18 ^{+ 2.73}_{- 4.12}}$  & {0.33}& ${ 5.69 ^{+ 0.58 }_{-0.11 }}$ & {0.09} & ${ 5.72 ^{+ 0.47 }_{-0.32 }}$ & {0.02} & ${ 11.31 ^{+ 0.63 }_{- 0.64 }}$&{0.04}  & ${ 11.27 ^{+ 0.71 }_{- 0.65 }}$&{0.01}  & ${ 7.62 ^{+ 1.84 }_{- 1.34 }}$&{0.69}  & ${ 7.25 ^{+ 1.54 }_{- 1.11 }}$ &{0.18}\\[0.1cm]
    
    W & ${ 8.32 ^{+ 2.43 }_{- 5.22 }}$&{0.79}  & ${ 5.76 ^{+ 2.62 }_{- 3.97 }}$& {0.17}  & ${ 5.86 ^{+ 0.51 }_{- 0.33 }}$ & {0.03} & ${ 5.69 ^{+ 0.48 }_{-0.34 }}$ & {0.00}  & ${ 11.33 ^{+ 0.65 }_{- 0.81 }}$&{0.02}  & ${ 11.24 ^{+ 0.64 }_{-0.58 }}$&{0.01}  & ${ 6.46 ^{+ 1.73 }_{- 1.24 }}$&{0.23}  & ${ 7.28 ^{+ 1.32 }_{- 1.19 }}$ &{0.06} \\[0.1cm]
    
    S & ${ 9.91^{+ 2.92}_{- 2.12}}$&{0.11}  & ${ 5.62 ^{+ 3.14}_{- 3.93}}$& {0.24}  & ${ 5.84 ^{+ 0.75 }_{-0.63 }}$ & {0.03}  & ${ 5.68 ^{+ 0.46 }_{-0.33 }}$ & {0.01}  & ${ 11.28 ^{+ 0.93 }_{- 0.64 }}$&{0.02}  & ${ 11.20 ^{+ 0.56 }_{- 0.52 }}$&{0.01}  & ${ 6.25 ^{+ 1.38 }_{- 1.30 }}$&{0.62}  & ${ 7.40 ^{+ 1.39 }_{- 1.31 }}$ &{0.09}\\[0.1cm]
    \bottomrule

    In-V & ${ 6.56 ^{+ 2.66 }_{- 1.56 }}$&{1.19}  & ${ 5.41 ^{+ 2.53}_{- 3.47 }}$ &{0.45} & ${ 6.27 ^{+ 0.33 }_{- 0.23 }}$&{0.18}  & ${ 5.93 ^{+ 0.44 }_{- 0.32 }}$ &{0.03} & ${ 11.63 ^{+ 0.66 }_{- 0.80 }}$ &{0.04} & ${ 11.48 ^{+ 0.67 }_{- 0.59 }}$ &{0.01} & ${ 6.87 ^{+ 1.18 }_{- 1.46 }}$ &{0.98}  & ${ 7.36 ^{+ 1.22 }_{- 0.99 }}$&{0.19} \\[0.1cm]
    Out-V & ${ 6.78 ^{+ 2.39 }_{- 1.7 }}$ &{1.12} & ${ 5.56 ^{+ 2.67 }_{- 3.39}}$ &{0.50} & ${ 6.21 ^{+ 0.34 }_{- 0.09 }}$ &{0.18} & ${ 6.02 ^{+ 0.40 }_{- 0.24 }}$ &{0.05} & ${ 11.62 ^{+ 0.82 }_{- 0.80 }}$ &{0.03}  & ${ 11.55 ^{+ 0.77 }_{- 0.76 }}$&{0.01}  & ${ 6.92 ^{+ 1.27 }_{- 1.33 }}$ &{0.62} & ${ 7.35 ^{+ 0.99 }_{- 0.75 }}$ &{0.17}\\[0.1cm]
    W & ${ 6.6 ^{+ 2.59}_{- 4.09 }}$ &{0.57} & ${ 6.21 ^{+ 2.42 }_{- 2.66 }}$ &{0.15} & ${ 6.26 ^{+ 0.45 }_{- 0.47 }}$ &{0.03} & ${ 5.94 ^{+ 0.47 }_{- 0.25 }}$ &{0.01}  & ${ 11.60 ^{+ 0.70 }_{- 0.54 }}$ &{0.02} & ${ 11.52 ^{+ 0.68 }_{- 0.59 }}$ &{0.01} & ${ 6.63 ^{+ 1.54 }_{- 1.19 }}$ &{0.31} & ${ 7.35 ^{+ 1.13 }_{- 0.97 }}$ &{0.08} \\[0.1cm]
    S & ${ 6.56 ^{+ 2.66 }_{- 1.56}}$ &{0.80} & ${ 5.44^{+ 2.53 }_{- 3.47}}$ &{0.24} & ${ 6.27 ^{+ 0.33 }_{- 0.23 }}$&{0.10}  & ${ 5.93 ^{+ 0.44 }_{- 0.32 }}$ &{0.02}  & ${ 11.63 ^{+ 0.66 }_{- 0.80 }}$ &{0.03} & ${ 11.48 ^{+ 0.67 }_{- 0.59 }}$ &{0.01} & ${ 6.87 ^{+ 1.18 }_{- 1.46 }}$ &{0.31} & ${ 7.36 ^{+ 1.22 }_{- 0.99 }}$&{0.12}\\[0.1cm]
    \bottomrule

    In-V & ${ 3.87 ^{+ 1.42 }_{- 2.12 }}$ &{1.18} & ${ 3.31 ^{+ 2.63 }_{- 2.29}}$ &{1.04} & ${ 6.89 ^{+ 0.55 }_{- 0.43 }}$ &{0.23} & ${ 6.97 ^{+ 0.25 }_{- 0.19 }}$ &{0.15} & ${ 12.12 ^{+ 0.52 }_{- 0.55 }}$ &{0.07} & ${ 11.98 ^{+ 0.58 }_{- 0.55 }}$ &{0.05} & ${ 7.91 ^{+ 0.91 }_{- 0.74 }}$ &{0.59} & ${ 8.48 ^{+ 1.18 }_{- 0.89 }}$ &{0.31}\\[0.1cm]

    Out-V & ${ 4.34 ^{+ 1.84 }_{- 3.16 }}$ &{1.43} & ${ 3.77 ^{+ 1.93 }_{- 4.99 }}$ &{1.56} & ${ 6.82 ^{+ 0.52 }_{- 0.01 }}$ &{0.16} & ${ 6.64 ^{+ 0.31 }_{- 0.21 }}$ &{0.11} & ${ 11.99 ^{+ 0.68 }_{- 1.20 }}$ &{0.04} & ${ 11.81 ^{+ 0.79 }_{- 0.26 }}$ &{0.07} & ${ 7.71 ^{+ 1.37 }_{- 0.47 }}$ &{0.57} & ${ 8.10 ^{+ 1.44 }_{- 1.00 }}$&{0.33} \\[0.1cm]
    
    W & ${ 5.05 ^{+ 1.61 }_{- 1.76 }}$ &{0.51} & ${ 4.71 ^{+ 2.15}_{- 2.91 }}$ &{0.35} & ${ 6.95 ^{+ 0.29 }_{- 0.06 }}$ &{0.10} & ${ 6.52 ^{+ 0.34 }_{- 0.48 }}$ &{0.06} & ${ 11.97 ^{+ 0.64 }_{- 0.19 }}$ &{0.05} & ${ 11.89 ^{+ 0.66 }_{- 0.30 }}$ &{0.02} & ${ 7.57 ^{+ 1.71 }_{- 1.10 }}$ &{0.29} & ${ 7.68 ^{+ 1.01 }_{- 0.87 }}$&{0.12}\\[0.1cm]
    
    S & ${ 4.52 ^{+ 3.53 }_{- 1.66 }}$ &{1.53} & ${ 5.01 ^{+ 3.17 }_{- 2.53}}$ &{0.44} & ${ 6.84 ^{+ 0.27 }_{- 0.22 }}$ &{0.16} & ${ 6.44 ^{+ 0.35 }_{- 0.25 }}$ &{0.05} & ${ 11.98 ^{+ 0.51 }_{- 0.63 }}$ &{0.04} & ${ 11.85 ^{+ 0.71 }_{- 0.35 }}$ &{0.02} & ${ 7.41 ^{+ 1.94 }_{- 1.39 }}$ &{0.57} & ${ 7.59 ^{+ 0.87 }_{- 0.86 }}$ &{0.15} \\[0.1cm]
    \bottomrule
    
  \end{tabular}
  }
  \tablefoot{The median sSFR, $\mbh$, $\mhalo$ and $\tlau(\mstar)$ and $\tlau$($\mhalo$) for AGN and non-AGN galaxies in  the four defined different cosmic environments and in the three different \\stellar mass intervals:  9.0 < log($M_{*}$) < 9.5 (top panel),  9.5 < log($M_{*}$) < 10.0 (middle panel) and  log($M_{*}$) > 10.0 (bottom panel). We  present the medians and  the semi-interquartile\\ range in the first column, where the uncertainty corresponds to the difference between the median and the $25-75^{th}$ percentiles {and in the second column we present the standard \\deviation of the bootstrap medians.}}
  \label{tab:galparams}

\end{sidewaystable*}

\end{appendix}
\end{document}